\documentclass[aip,pof,preprint]{revtex4-1}
\usepackage{physics,comment,float,subcaption,amsmath,amssymb}
\usepackage{graphicx,hyperref}
\graphicspath{{images/}}
\draft % marks overfull lines with a black rule on the right

\begin{document}

%\preprint{}

\title{A numerical study on the influence of geometry on the rupture risk of abdominal aortic aneurysms} %Title of paper



\author{G R Krishna Chand Avatar}
\email[]{krishnaagr@iisc.ac.in}
%\thanks{}
\affiliation{Department of Aerospace Engineering, Indian Institute of Science, Bangalore, India}

\author{Chinika Dangi}
\email[]{chinikadangi@iisc.ac.in}

\author{Puneet Pushkar}
\email[]{puneetp@alum.iisc.ac.in}


%\date{\today}

\begin{abstract}
Abdominal aortic aneurysms (AAAs) are local dilatations in the abdominal aorta
occurring due to weakening of arterial wall. The present work investigates the influence of the ratio of maximum transverse diameter to abdominal height (DHr) on rupture risk of AAA, using  hemodynamics and AAA wall mechanics simulations. We have considered two idealized AAA geometries, AAA1 of higher DHr than AAA2. Two constitutive models, namely, Newtonian and Carreau-Yasuda models have been used for modelling blood as an incompressible fluid. Additionally, in order to describe the behaviour of AAA wall, three constitutive models, namely, linear elastic, Saint Venant Kirchhoff elastic and a phenomenological finite-strain model called Raghavan-Vorp elastic have been utilised. Numerical simulation of AAA biomechanics have been performed using \textit{solids4Foam}, an  open source package built on finite volume framework. Hemodynamic parameteric study reveals that AAA1 has lower time-averaged wall shear stress (TAWSS) and higher oscillatory shear index (OSI) compared to AAA2. Thus, AAA1 has increased susceptibility to thrombus deposition and, therefore, higher risk of AAA rupture. Furthermore, the peak wall stress of AAA1 is about 8\% higher than that of AAA2.  It is concluded that higher DHr leads to greater rupture risk.

%Raghavan-Vorp model produces the highest peak wall stress than the other two models.

\end{abstract}


\maketitle %\maketitle must follow title, authors, abstract and \pacs

\section{Introduction}
\begin{figure}[h!]
	\centering
	\includegraphics[width=0.35\linewidth]{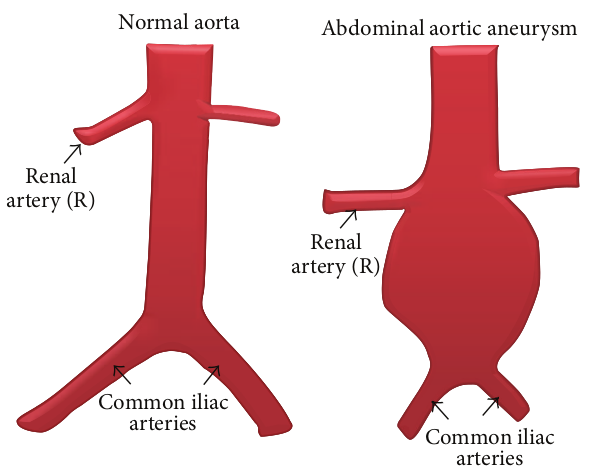}
	\caption{Schematic of normal aorta and abdominal aortic aneurysm. Source: \cite{Canchi2015}}
	\label{fig:abdominal-aortic-aneurysm}
\end{figure}

An arterial aneurysm is a vascular condition characterized by an abnormal focal dilatation of an artery with respect to its original state \cite{Canchi2015,Aggarwal2011,Philip2022}. This happens due to localised bulging of blood vessels as a result of the weakening of the arterial walls \cite{Philip2020}. Aneurysm shapes could be saccular (asymmetric balloon-like expansions of a portion of the wall), fusiform (axisymmetric gradual dilatation of the entire circumference of the artery) or cylindroid \cite{Finol2003}. Aortic aneurysms are located either in the thoraric region or in the infrarenal segment, that is, the abdominal region between the renal arteries and the iliac bifurcation; the latter is called abdominal aortic aneurysm (AAA) (Fig. \ref{fig:abdominal-aortic-aneurysm}) and is nearly 80\% of all aortic aneurysms. There are several risk factors responsible for the genesis, growth and rupture of AAAs, namely, people aged above 60 years, smoking, males, patients with artherosclerosis, coronary heart disease and hypertension \cite{Ullery2018,Lederle2000,Lederle1997}. AAAs are severe medical conditions often requiring surgical management. AAAs are often asymptomatic until they suddenly rupture, which necessitates a measure of rupture risk prediction. Typically, surgeons tend to use a criterion based on maximum transverse diameter to decide on the course of treatment. Fusiform AAAs with diameter larger than 5-5.5 cm are recommended for surgical repair while those smaller than 4 cm are considered to be at low risk of rupture \cite{Raut2013}. However, this criterion has been deemed crude by various researchers \cite{Vorp1998,Finol2001,Finol2003,Finol2003a,Raut2013,Boyd2016} since there have been instances when aneurysms of diameter smaller than 4 cm ruptured while those with diameter greater than the critical transverse diameter did not \cite{Venkatasubramaniam2004}; thus necessitating detailed hemodynamics and wall mechanics studies to identify the parameters directly and indirectly responsible for the rupture of AAAs\cite{Mutlu2023}. These parameters include wall shear stress and its derivatives, and aneurysmal wall stress.

%\subsection{Role of numerical simulations for the study of aneurysms}

Bluestein et al. \cite{Bluestein1996} performed computational fluid dynamics (CFD) simulation of an axisymmetric AAA and deduced that flow recirculation regions creates suitable conditions encouraging thrombus formation and the viability of aneurysm rupture. Finol et al. \cite{Finol2003a} did CFD simulations of patient-specific aneurysm models to quantify vortex dynamics and flow-induced  wall shear stress (WSS) distributions and found that the vortex dynamics is specific to every individual and is dependent on the shape and size of the aneurysm. Boyd et al. \cite{Boyd2016} conducted a CFD study on ruptured AAAs and observed that aortic rupture took place at the sites of low WSS. Philip et al. \cite{Philip2020} postulated that high time-averaged wall shear stress (TAWSS) and low oscillatory shear index (OSI) indicates the initiation of intraluminal thrombus (ILT) formation. Furthermore, Vorp et al. \cite{Vorp1998} conducted a numerical study on AAAs and found that aneurysm rupture occurs when the mechanical stress acting on the aneurysm inner wall due to intraluminal pressure, exceeds the failure strength of the degenerated aortic tissue. Raghavan et al. \cite{Raghavan1996} conducted an experimental study on 78 tissue samples (71 with AAAs), and observed that yield strength reduced by nearly 50\% and ultimate strength reduced by over 50\% for AAAs with respect to the normal aorta. 
Venkatasubramaniam et al. \cite{Venkatasubramaniam2004} found that peak wall stress was significantly higher in the ruptured AAA and also discovered that the area of peak wall stress correlated with rupture site. 

Martufi et al. \cite{Martufi2009} studied 3D axisymmetric AAA and observed that DHr (ratio of maximum transverse diameter $D_{max}$ to abdominal height $H$), DDr (ratio of maximum transverse diameter to proximal neck diameter), Hr (ratio of abdominal height to neck height) and BL (ratio of bulge height to abdominal height) are important shape parameters in assessing the aneurysm risk prediction. However, few researchers\cite{Cappeller1997,Perez2016,Philip2020} have suggested that DHr and DDr could be more accurate parameters and requires further detailed analysis. Recently, Kumar et al. \cite{Kumar2023} investigated the influence of AAA shape and deduced that maximum dilation diameter and aneurysm neck angle affects local hemodynamic parameters considerably. Additionally, Faraji et al. \cite{Faraji2022} studied the role of four different viscosity models on pulsatile blood flow of AAA and concluded that non-Newtonian model predicted OSI more accurately.  Dalbosco et al. \cite{Dalbosco2023} conducted a multi-scale computational study on AAA and correlated the growth of AAA with alterations at microstructural level in the aortic tissue.

To the best of authors' knowledge, no parametric study has been reported in literature on AAA biomechanics simulations based on Raghavan-Vorp model. The present work is a complete study to investigate the influence of DHr on both hemodynamics and arterial wall mechanics using realistic constitutive models. We have modeled blood using both Newtonian and  Carreau-Yasuda (non-Newtonian) constitutive relations. The AAA wall has been modeled using both linear elastic and non-linear elastic (Saint Venant Kirchhoff elastic and Raghavan-Vorp elastic \cite{Raghavan2000}) models. We consider two AAA geometries, AAA1 and AAA2, whose shape indices are given in table below \cite{Philip2020}
\begin{table}[h!]
	\begin{center}
		\begin{tabular}{|c|c|c|}
			\hline
			& AAA1  &  AAA2    \\
			\hline
			DHr  &  0.83    & 0.45  \\
			\hline
			DDr  & 2   & 2    \\
			\hline
			Hr &   2.4   &  2.4 \\ % 2.4, 3.67
			\hline
			BL &  0.92   &  0.92 \\ % 0.92, 0.77  
			\hline
		\end{tabular}
	\end{center}
\end{table}.
Hemodynamics and computational solid stress simulations has been performed using \textit{solids4Foam}, an OpenFOAM-based open source package for solving continnum mechanics problems in finite volume framework. Computational models of AAA lumen geometry were prepared using CAD software \textit{Salome}. A numerical study investigating the effect of blood constitutive models, effect of different shape indices (AAA1 and AAA2), influence of AAA wall contitutive model and DHr has been presented. The analysis highlights importance of geometry and shape of AAAs on rupture risk.

%% METHODOLOGY
\section{Problem formulation}  \label{chap:methodology} 
%\section{Methodology}

\begin{figure}[h!]
	\centering
	\includegraphics[width=0.8\linewidth]{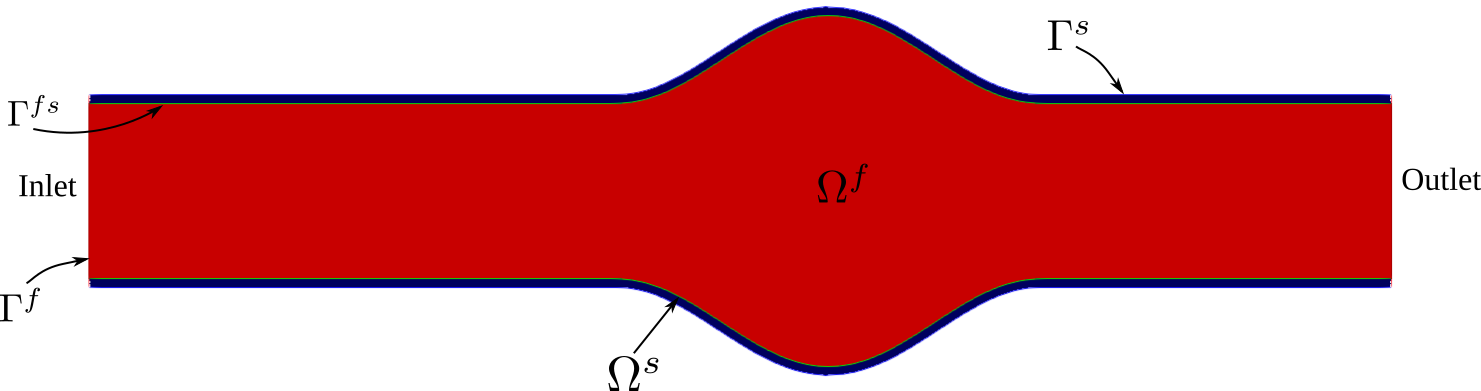}
	\caption{Fluid domain $\Omega^f$ with boundary $\Gamma^f$ enclosing it, solid domain $\Omega^s$ with its corresponding boundary $\Gamma^s$, and fluid-solid interface $\Gamma^{fs}$}
	\label{fig:domain}
\end{figure}

%\subsection{Physical and Mathematical models}
%Physically, the aneurysm problem is composed of two sub-parts, namely, fluid part and solid part. The fluid part involves computation of the flow dynamics of blood through the AAA lumen (fluid domain) while the solid part involves computing displacement of the elastic arterial and aneurysm wall (solid domain) due to traction exerted on its endoluminal surface by the blood flow. 
The geometry of the aneurysm problem is denoted by a domain $\Omega$, with boundary $\Gamma$, which is composed of a solid domain $\Omega^s$ and a fluid domain $\Omega^f$, such that $\Omega = \Omega^s \cup \Omega^f$, alongwith their corresponding boundaries $\Gamma^s$ and $\Gamma^f$  as shown in Fig. \ref{fig:domain}. The fluid-solid interface is $\Gamma^{fs} = \Gamma^f \cap \Gamma^s$ which acts as the wall for fluid domain). 

%The problems in continuum mechanics are governed by the laws of conservation of mass, momentum and energy. These laws are mathematically represented as a set of partial differential equations. 

\subsection{Fluid dynamical model for Blood}
We consider blood to be an incompressible fluid which shows both shear-thinning and viscoelastic behaviour. Sequeira et al. \cite{Sequeira2018} have stated that blood flow displays predominantly Newtonian behaviour as it circulates through large arteries. However, Khanafer et al. \cite{Khanafer2006} have concluded that the blood flow may be affected by non-Newtonian behaviour of blood in pathological vascular conditions such as AAAs where RBCs and platelets can aggregate resulting in the change of viscosity. Therefore, we carry out numerical simulations using both Newtonian and non-Newtonian Carreau-Yasuda models for the blood with material parameters taken from Biasetti et al \cite{Biasetti2011}.
% Non-Newtonian effects in AAA paper 
The blood density is $\rho = 1050$ kg/m$^3$ and the transport properties for the two models are 
\begin{itemize}
	\item Newtonian model: dynamic viscosity $\mu^f = 0.0044$ Pa-s 
	
	\item Carreau-Yasuda model:  dynamic viscosity $\mu^f = \mu_\infty + (\mu_0 - \mu_\infty)[ 1 + (\lambda \dot{\gamma})^a]^{(n-1)/a}$ , where
	$\mu_0 = 0.16$ Pa-s, $\mu_\infty = 0.0035$ Pa-s, $\lambda = 8.2$ s, $n = 0.2128$, $a = 0.64$
\end{itemize} 

%\subsubsection{Mathematical model}
The governing equations (incompressible Navier-Stokes equations) are derived from the conservation of mass and balance of linear momentum (neglecting body forces) and expressed in integral form as
\begin{equation}
	\begin{split}
		&\int_{S} \rho^f \vb*{v}^f \cdot \vb*{n} dS = 0 \\
		&\pdv{}{t}\int_{V} \rho^f \vb*{v}^f (\vb*{x},t) dV + \oint_{S} \rho^f \vb*{v}^f\vb*{v}^f\cdot \vb*{n} dS = \oint_{S} \vb*{\sigma}^f\cdot \vb*{n}dS 
	\end{split}
    \label{eqn:fluid-governing-equations}
\end{equation}
The consitutive relation relating the Cauchy stress tensor to the velocity field for a Newtonian fluid is given as $\vb*{\sigma}^f = -p\vb*{I} + \mu^f \bigg(\grad{\vb*{v}^f} + (\grad{\vb*{v}^f})^T\bigg) - \frac{2}{3}\mu^f (\div{\vb*{v}^f})\vb*{I}$, where $\vb*{I}$ is the second order identity tensor, $p$ is the pressure field.

The integral equations (Eqn. \ref{eqn:fluid-governing-equations}) are discretized using finite volume method. 
The resulting discrete system is solved using Pressure Implicit with Splitting of Operators (PISO) algorithm in a segregated fashion. An unconditionally stable second-order accurate backward Euler scheme is used to advance the simulation. The linear systems for velocity components are solved using preconditioned bi-conjugate gradient (PBiCG) method with tolerance of $1 \times 10^{-8}$ while matrix equation for pressure is solved using geometric-agglomerated algebraic multi-grid (GAMG) solver with a tolerance of $1 \times 10^{-6}$.

%Upon substituting \ref{eqn:constitutive-relation-Newtonian-fluid} in \ref{eqn:momentum-balance-eqn} and using continuity equation, we have
%\begin{align}
%\pdv{}{t}\int_{V(t)} \rho^f \vb*{v}^f (\vb*{x},t) dV + \oint_{S(t)} \rho^f \vb*{v}^f\vb*{c}\cdot \vb*{n} dS &= - \oint_{S(t)} p\vb*{n}dS + \oint_{S(t)} \nu^f \grad{\vb*{v}^f}\cdot \vb*{n} dS \label{eqn:momentum-balance-eqn-newtonian-fluid}
%\end{align}

%\subsubsection{Fluid mesh motion}
%The fluid mesh has to updated according to the movement of the solid domain (aneurym). In order to avoid large distortion of the cells, a dynamic mesh methodology is used. A technique based on the Laplace equation's solution for the velocity of cell centroids, also referred to as Laplace smoothing, is utilized for this purpose which involves solving the following equation:
%\begin{align}
%\div{\gamma \grad{\vb*{\omega}}} &= 0 \label{eqn:fluid-mesh-motion}
%\end{align} 
%where $\gamma$ is a ``mesh diffusivity'' \cite{Jasak2006} and $\vb*{\omega}$ is the control volume centroid velocity.

\begin{figure}[h!]
	\centering
	\includegraphics[width=\linewidth]{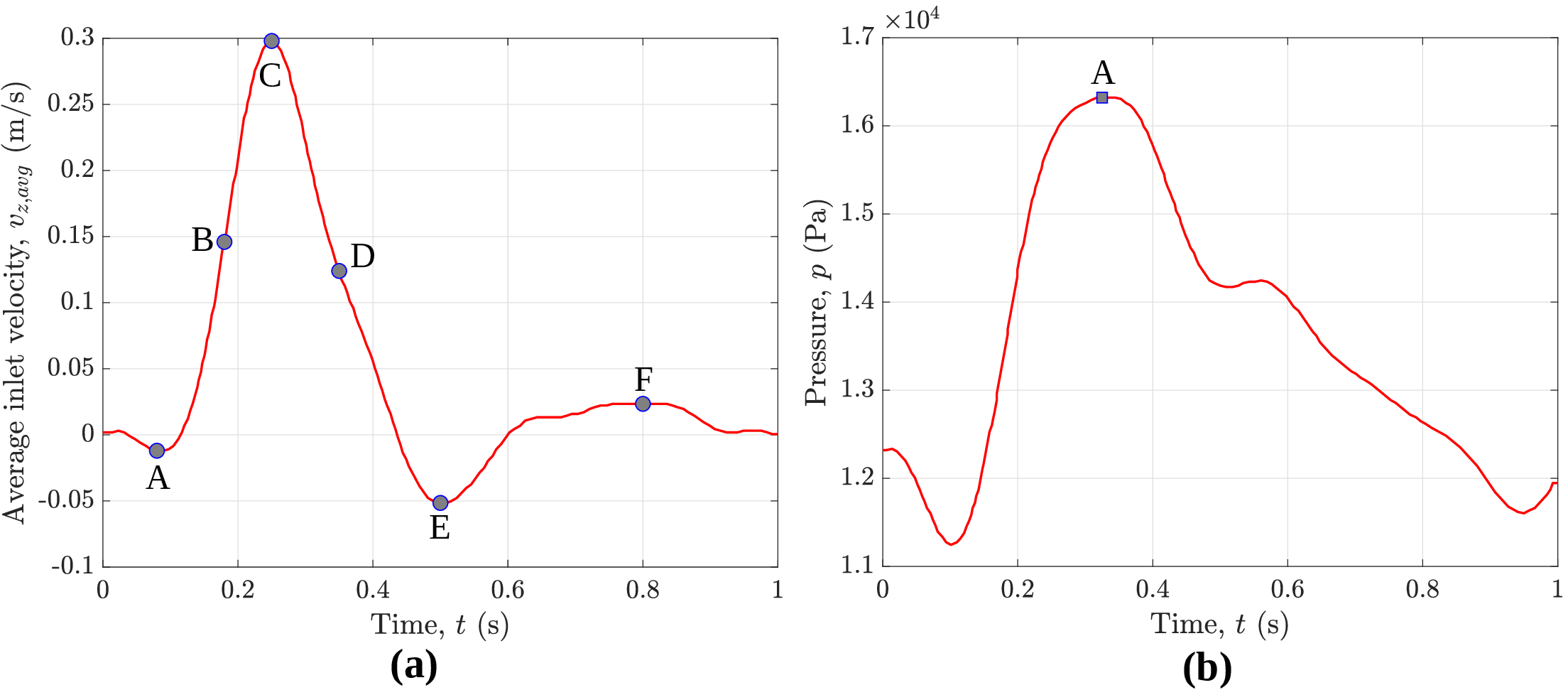}
	\caption{Waveforms for (a) Average inlet velocity with markers indicating time instants: initiation of systole A (t = 0.08 s), systolic acceleration B (t = 0.18 s), peak systole C (t = 0.25 s), systolic deceleration D (t = 0.35 s), beginning of diastole E (t = 0.50 s), late diastole F (t = 0.80 s)  (b) Outlet pressure with marker A (t = 0.32 s) indicating peak pressure. Source: adapted from \cite{Philip2020}}
	\label{fig:velocity-pressure-waveforms}
\end{figure}

%Upon solving eqn \ref{eqn:fluid-mesh-motion} subject to the above boundary conditions, the cell centroid velocity $\vb*{\omega}$ is interpolated from the cell centroid to the geometric nodes which we call $\vb*{\omega}_{node}$ \cite{Oliveira2017}. Finally, we update the mesh node coordinates by
%\begin{align}
%\vb*{x}_{new} &= \vb*{x}_{old} + \vb*{\omega}_{node} \Delta t
%\end{align}
%where $\Delta t$ is the time-step of the numerical simulation, $\vb*{x}_{new}$ and $\vb*{x}_{old}$ are the mesh nodes position after and before solving the equation, respectively.
\paragraph{Initial and Boundary conditions} We set uniform values of velocity and pressure field throughout the domain as initial conditions. The boundary conditions are:
\begin{itemize}
	\item Inlet: A time-varying paraboloid velocity profile whose cross-sectional average value is shown in Fig. \ref{fig:velocity-pressure-waveforms}. We use an 8-series Fourier decomposition for the average velocity. Zero-gradient condition is used for pressure. %The paraboloid velocity profile is given by $v_z (r, t) = 2 v_{z,avg} (t)  \bigg(1 - \left(\frac{r}{R}\right)^2\bigg)$  where $R$ is the radius of the inlet face, $r$ is the radial distance from the centroid of the face. 
	
	\item Outlet: We impose a time-varying pressure as shown in Fig. \ref{fig:velocity-pressure-waveforms} via 8-series Fourier decomposition for pressure. Zero-gradient condition is used for velocity. % Flux-corrected velocity
	
	\item Endoluminal surface of AAA wall: No-slip condition is prescribed for velocity such that $\vb*{v} = \vb*{0}$ while zero-gradient is used for the pressure field. 
\end{itemize}

%\begin{figure}[h!]%
%	\centering
%	\subfloat[\centering Inlet velocity waveform]{{\includegraphics[width=0.45\textwidth]{philip-inlet-velocity} } \label{fig:philip-inlet-velocity}}%
%	\qquad
%	\subfloat[\centering Outlet pressure waveform]{{\includegraphics[width=0.45\textwidth]{philip-outlet-pressure} } \label{fig:philip-outlet-pressure}}%
%	\caption{Pressure and velocity waveforms}%
%	\label{fig:example}%
%\end{figure}

\subsection{Solid dynamical model for the Aneurysm wall}
%\subsubsection{Physical model}
We model the AAA wall as linear elastic \cite{Scotti2005,Philip2020} and Saint Venant Kirchhoff elastic \cite{Scotti2005}. We also employ a non-linear isotropic phenomenological constitutive model for AAAs proposed by Raghavan et al. \cite{Raghavan2000}; we refer to this as Raghavan-Vorp elastic model. The mechanical properties of the AAA wall are assumed to be homogeneous; density $\rho = 2000$ kg/m$^3$ and material constants are 
\begin{itemize}
	\item Linear and Saint Venant Kirchhoff elastic models: $E = 2.7 \times 10^6$ Pa, $\nu^s = 0.45$ 
	\item Raghavan-Vorp model:, $\alpha = 1.74 \times 10^5$ Pa, $\beta = 1.881 \times 10^6$ Pa, $\nu^s = 0.45$
\end{itemize}
The wall thickness is considered to be uniform with a value of $1.2$ mm \cite{Philip2020}.

\begin{comment}
	Consider a generic solid body (fig \ref{fig:generic-solid-body}) with volume $\Omega$ and surface $\Gamma$ where some part of its boundary is subjected to a prescribed displacement field $\vb*{u}_b$ and the remainder is provided a traction force $\vb*{T}_b$.
	
	\begin{figure}[h!]
		\centering
		\includegraphics[width=0.5\linewidth]{generic_solid_body}
		\caption{Generic solid boundary. Source \cite{Cardiff2021}}
		\label{fig:generic-solid-body}
	\end{figure}
	
	{\bfseries Explain Poisson's ratio}
\end{comment}

%\subsubsection{Mathematical model}
We use total Lagrangian formulation \cite{Belytschko2014,Holzapfel2000nonlinear} of the governing equations for solid mechanics, where deformations and stresses are calculated with respect to the reference or initial configuration. We represent the deformation by the deformation gradient tensor $\vb*{F}$  and the stress by the first Piola-Kirchhoff stress tensor $\vb*{P}$. $\vb*{F}$ is expressed as function of the solid displacement $\vb*{u}$ as $\vb*{F} = \vb*{I} + \grad_0{\vb*{u}}$
where subscript $0$ indicates differentiation with respect to the intial configuration. The balance of linear momentum for a deformable solid, neglecting body force, is
\begin{align}
	\rho_0^s \pdv[2]{\vb*{u}}{t} &= \grad_0 \cdot \vb*{P}  \label{eqn:general-balance-of-linear-momentum}
\end{align}

The first and second Piola-Kirchhoff stress tensors \cite{Holzapfel2000nonlinear}, denoted as $\vb*{P}$ and  $\vb*{S}$ respectively, are related to each-other as $\vb*{P} = \vb*{F}\cdot \vb*{S}$
and to Cauchy stress tensor $\vb*{\sigma}$ as 
\begin{align}
	\vb*{\sigma} &= \frac{1}{J} \vb*{P} \cdot \vb*{F}^T = \frac{1}{J} \vb*{F}\cdot \vb*{S} \cdot  \vb*{F}^T  \label{eqn:relation-between-cauchy-piola-kirchhoff-tensors}
\end{align}
where $J = \det(\vb*{F})$. The Green-Lagrange strain tensor is defined as 
\begin{align}
	\vb*{E} &= \frac{1}{2}\left(\vb*{F}^T\cdot \vb*{F} - \vb*{I}\right) = 
	\frac{1}{2}\big(\underbrace{\grad_0{\vb*{u}} + (\grad_0{\vb*{u}})^T}_{\text{linear part}} + \underbrace{(\grad_0{\vb*{u}})^T\cdot \grad_0{\vb*{u}}}_{\text{non-linear part}}\big) \label{eqn:green-lagrange-strain-tensor}
\end{align}
Now, $\vb*{E}$ is composed of linear as well as non-linear parts making it appropriate for measuring large deformations in elastic solids which shows a non-linear behaviour. A non-linear elastic model used to capture small strains and finite rotations is the Saint Venant-Kirchhoff model. For an istropic material, the constitutive relation is given as 
\begin{align}
	\vb*{S} &= 2 \mu^s \vb*{E} + \lambda^s \tr(\vb*{E}) \vb{I} \label{eqn:st-venant-kirchhoff-elastic-model}
\end{align}
where $\mu^s$ (also called shear modulus \cite{Chaves2013}) and $\lambda^s$  are Lam\'{e} constants which are related to Young's modulus and Poisson ratio as \cite{Jasak2000} 
	$\mu^s = \frac{E}{2(1 + \nu^s)}  \text{ and } \lambda^s = \frac{\nu^s E}{(1 + \nu^s)(1 - 2\nu^s)}$

\paragraph{Linear elastic model: } Cauchy infinitesimal strain tensor \cite{Fung1990} defined as $\vb*{\epsilon} = \frac{1}{2}\big(\grad_0{\vb*{u}} + (\grad_0{\vb*{u}})^T)$
Also, $\vb*{S}$ reduces to Cauchy stress tensor $\vb*{\sigma}$. Thus constitutive relation is the isotropic Hooke's law 
\begin{align}
	\vb*{\sigma}^s &= 2 \mu^s \vb*{\epsilon} + \lambda^s \tr(\vb*{\epsilon}) \vb*{I}
\end{align}

\paragraph{Raghavan-Vorp elastic model} Raghavan et al. \cite{Raghavan2000} obtained the following constitutive equation relating the Cauchy stress tensor and the elastic deformation
\begin{align}
	\vb*{\sigma}^s &= - p\vb*{I} + 2 \left[\alpha + 2 \beta (I_{\vb*{b}} - 3)\right] \vb*{b} \label{eqn:incompressible-raghavan-vorp-elastic-model}
\end{align}
where $p$ is the Lagrange multiplier to enforce incompressibility, $\alpha$ and $\beta$ are material constants obtained after fitting uniaxial loading experimental data to the proposed model, $\vb*{b} = \vb*{FF}^T$ is left Cauchy-Green tensor and $I_{\vb*{b}}$ is its first principal invariant. Since arterial wall is nearly (but not exactly) incompressible, we developed a weakly compressible version of Raghavan-Vorp model for which the constitutive relation is given as
\begin{align}
	\vb*{\sigma}^s &= \frac{\kappa}{2}\bigg(J - \frac{1}{J}\bigg)\vb*{I} + \text{dev}\bigg(\frac{2}{J} \bigg[\alpha + 2\beta (I_{\overline{\vb*{b}}} - 3)\bigg] \overline{\vb*{b}}\bigg) \label{eqn:compressible-raghavan-vorp-elastic-model}
\end{align}
where $\overline{\vb*{b}} = J^{-2/3} \vb*{b}$ and $\kappa = \frac{2}{3}\mu^s + \lambda^s $

From Eqn. \ref{eqn:relation-between-cauchy-piola-kirchhoff-tensors}, we have 
\begin{align}
	\vb*{\sigma}^s &= \frac{1}{J} \vb*{P} \cdot \vb*{F}^T \implies \qquad \vb*{P} = J \vb*{\sigma}^s \vb*{F}^{-T} \label{eqn:relation-between-P-and-sigma}
\end{align}
Using Eqns. \ref{eqn:relation-between-P-and-sigma} and \ref{eqn:compressible-raghavan-vorp-elastic-model} in Eqn. \ref{eqn:general-balance-of-linear-momentum}, we obtain the governing equation for Raghavan-Vorp elastic material which is solved to obtain the solid displacement field $\vb*{u}$.

The governing equations (Eqn. \ref{eqn:general-balance-of-linear-momentum}) for solid domain are discretized using finite volume method. The resultant discretized system of equations for components of displacement $\vb*{u}$ are solved using preconditioned conjugate gradient (PCG) method with tolerance of $1 \times 10^{-9}$. Second-order accurate backward Euler scheme is used for time integration. To enhance convergence, the displacement equations are under-relaxed.

\paragraph{Initial and Boundary conditions (BCs)} The governing equations for solid domain are solved for displacement field. We set the displacement and velocity fields to be zero throughout the domain as initial conditions. We impose zero displacement conditions,  $\vb*{u} = \vb*{0}$, at the extremities or side faces. The internal (endoluminal) surface is subjected to Neumann boundary conditions which is the time-varying pressure loading $p$ corresponding to the cardiac cycle (Fig. \ref{fig:velocity-pressure-waveforms}). The external surface is traction-free.

\subsection{Simulation setup}
We have the following expression for the idealized fusiform AAA lumen geometry in terms of the maximum AAA diameter $D_{max}$, proximal neck diameter $D_{neck2}$ and height of the aneurysm sac $H$ as 
\begin{align*}
	r(z) &= \bigg(\frac{D_{max} - D_{neck2}}{4}\bigg)\bigg[1 + \sin\bigg(\frac{2\pi z}{H} - \frac{\pi}{2}\bigg)\bigg] + \frac{D_{neck2}}{2} \quad; \quad  0 \leq z \leq H
\end{align*}
It is also assumed that the proximal and distal neck diameters are equal $D_{neck1} = D_{neck2}$. We take $D_{max} = 5$ cm.

\begin{figure}[h!]
	\centering
	\includegraphics[width=0.8\linewidth]{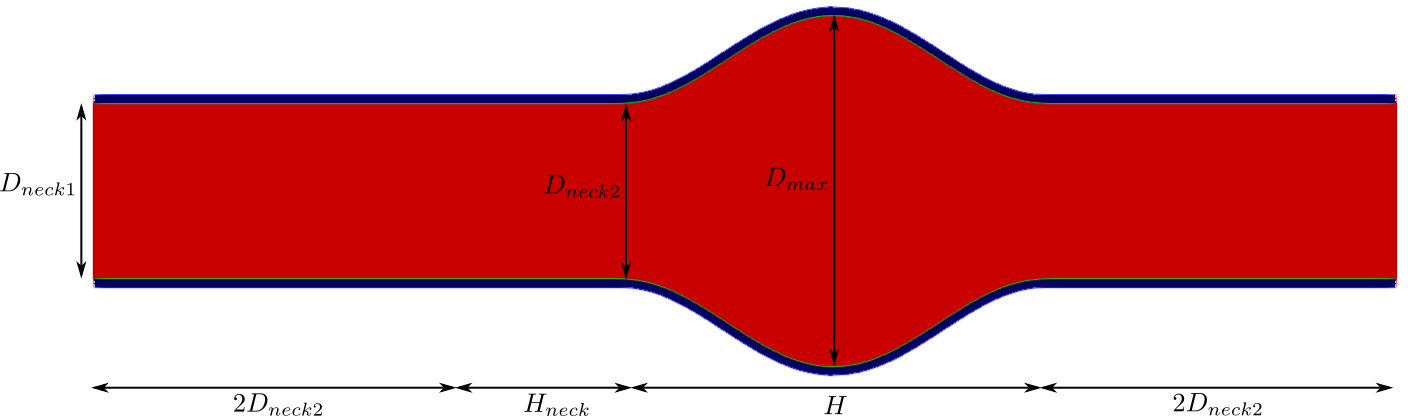}
	\caption{Idealized fusiform AAA geometry for simulation} % \cite{Philip2020}}
\label{fig:simulation-domain}
\end{figure}

\begin{figure}[h!]%
\centering
\subfloat[\centering Fluid mesh]{{\includegraphics[width=0.45\textwidth]{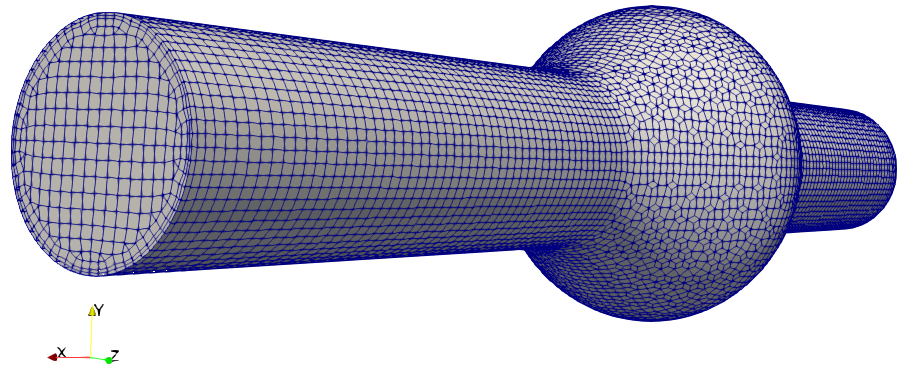} } \label{fig:simulation-domain-fluid-mesh}}%
\qquad
\subfloat[\centering Solid mesh]{{\includegraphics[width=0.45\textwidth]{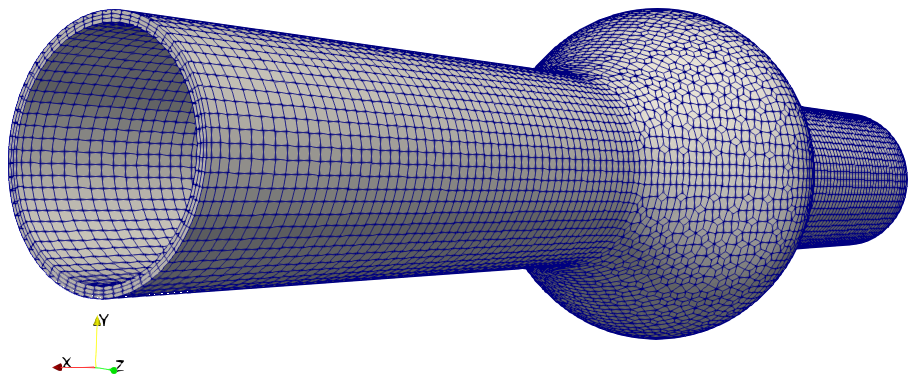} } \label{fig:simulation-domain-solid-mesh}}%
\caption{Computational mesh for (a) fluid and (b) solid domains}%
\label{fig:simulation-domain-mesh}%
\end{figure}

% $\int_{0}^T \abs{v_{z,avg}}dt \approx 6 \times 10^{-2}$ m or 6 cm
\paragraph{Simulation domain} For proper imposition of inlet and outlet boundary conditions, both inlet and outlet extensions are taken to be 2$D_{neck2}$ \cite{VamsiKrishna2020,Scotti2005,Philip2020}. We used an open-source computer-aided design (CAD) software \textit{Salome} \cite{Ribes2007} for preparing the AAA lumen geometries. The schematic of the AAA geometry is shown in Fig. \ref{fig:simulation-domain}. We used a fine mesh comprising of prismatic layers to properly resolve the unsteady boundary layer at the wall for fluid domain and capture flow dynamics accurately. As suggested in Krishna et al. \cite{VamsiKrishna2020}, boundary layer thickness $\delta$ for oscillatory flows is of the order {$\delta \sim \sqrt{\nu^f T}$}, where $\nu^f = \frac{\mu^f}{\rho^f}$ is the kinematic viscosity of the blood and $T$ is the time-period of the cardiac cycle. Using the blood properties, we obtain
\begin{align*}
	\delta \approx \sqrt{\frac{0.0044}{1050}\cdot 1} = 2 \times 10^{-3} \text{ m}
\end{align*}
implying that the fluid mesh ought to be refined upto a distance of $2 \times 10^{-3}$ m from the wall. 
An open-soure mesh generation tool called \textit{cfMesh} \cite{cfMesh} was used to generate fluid mesh for AAA lumen (Fig. \ref{fig:simulation-domain-fluid-mesh}). On the other hand, the solid mesh for each AAA geometry was generated by extruding the wall patch of the fluid mesh along the local wall-normal direction for AAA wall thickness, and number of layers (Fig. \ref{fig:simulation-domain-solid-mesh}).
We use an open-source package called \textit{solids4Foam} which provides a host of applications (solvers and utilities) to deal with a wide variety of problems in fluid dynamics, solid dynamics and fluid-structure interactions.

%\subsection{Hemodynamic and arterial wall stress parameters}

%%%%%%%%%% RESULTS
%\chapter{Results}%: Influence of AAA geometry on its rupture risk}
\section{Hemodynamics simulations of AAA lumen geometries}
We perform hemodynamic simulations using Newtonian and Carreau-Yasuda constitutive models for blood. 

For analyzing the blood flow in the AAAs, we consider the variation of wall shear stress (WSS) and its derived quantities. We also study vortex dynamics and identify the vortical (coherent) structures using $\lambda_2$-criterion\cite{Jeong1995,Zhang2023}, a velocity gradient-based vortex identification method.  Wall shear stress (WSS) vector $\vb*{\tau}_w$ is defined as the tangential component of the traction vector $\vb*{t}$ at the wall. Traction vector is obtained as $\vb*{t} = \vb*{\sigma}^f\cdot \vb*{n}^f$, where $\vb*{\sigma}^f$ is Cauchy stress tensor and $\vb*{n}^f$ is the wall normal unit vector. WSS vector \cite{OliveiraPrivate}  is obtained as $\vb*{\tau}_w = \vb*{t} - (\vb*{t}\cdot \vb*{n}^f) \vb*{n}^f$. We look at the following derived quantities to get a holistic view of the wall shear stress \cite{Browne2015,Oliveira2021,Philip2020,Mutlu2023}:
\begin{itemize}
	%	\item Mean wall shear stress ($meanWSS$): It is defined as the magnitude of temporal average of each cell-centered nodal WSS vector over a cardiac cycle of time period $T$,
	%	\begin{align*}
		%	meanWSS (\vb*{x}) &= \norm{\frac{1}{T}\int_0^T \vb*{\tau}_w (\vb*{x}, t)dt}
		%	\end{align*}
	
	\item Time-averaged wall shear stress ($TAWSS$) is calculated as $TAWSS (\vb*{x})  = \frac{1}{T}\int_0^T \norm{\vb*{\tau}_w (\vb*{x}, t)}dt$
	
	\item Oscillatory shear index ($OSI$) is a dimensionless metric which characterises whether the WSS vector is aligned with the $TAWSS$ vector throughout the cardiac cycle, $OSI (\vb*{x}) = \frac{1}{2} \bigg(1 -\frac{\norm{\frac{1}{T}\int_0^T \vb*{\tau}_w (\vb*{x}, t)dt}}{\frac{1}{T}\int_0^T \norm{\vb*{\tau}_w (\vb*{x}, t)}dt}\bigg)$
	
	\item Spatial gradient of the mean wall shear stress ($WSSG_S$) is calculated as $WSSG_S = \norm{\grad{(meanWSS)}}$
	
	\item Peak systole wall shear stress, $PSWSS (\vb*{x}) = \norm{\vb*{\tau}_w (\vb*{x}, t_{ps})}$ where $t_{ps}$ is the time instant of peak systole
	
	%	\item Instantaneous surface-averaged WSS ($SWSS$) on the aneurysm sac surface over time \cite{Oliveira2021}, 	$SWSS (t) = \frac{1}{A_a} \int_{A_a} \norm{\vb*{\tau}_w (\vb{x}, t)} ~dA_a$ where $A_a$ is the area of the arterial wall.	
\end{itemize}

At first, we carry out a comparative study of the effect of constitutive models on the hemodynamics within AAA1 for wall shear stress (WSS) and other derived quantities --  $PSWSS$, $meanWSS$, $WSSG_S$, $TAWSS$, $OSI$. Therefter we examine the role played by DHr on the flow dynamics within AAA1 and AAA2.

%\paragraph{Cardiac cycle solution convergence}
%Although the prescribed inlet and outlet boundary conditions for the fluid domain is periodic with a time-period $T = 1$ s, it takes a number of cycles for the numerical solution to achieve the same periodicity. Upon carrying out computations on the coarse mesh using the non-Newtonian Carreau-Yasuda constitutive model for 4 cardiac cycles, we found that the solutions corresponding to two consecutive cycles after the second cycle were indistinguishable. Therefore, we carry out the rest of the simulations for 4 cardiac cycles.  
% \paragraph{Mesh convergence study}

We carry out the simulations for 4 cardiac cycles and perform a mesh convergence study for AAA1 using three meshes with the number of cells increasing progressively with a factor of approximately 2 as tabulated in Table \ref{tab:mesh-convergence-AAA1}. The time-step size is also progressively decreased by a factor of 2 such that a constant maximum Courant-Friedrichs-Levy (CFL) number of 1, that is CFL $ \leq 1$, is maintained throughout every simulation.
\begin{table}[h!]
\begin{center}
\begin{tabular}{|c|c|c|c|}
	\hline
	&  Coarse   & Medium  &  Fine     \\
	\hline
	Number of cells (in millions) &  1.82   & 3.66   &  6.8 \\
	Number of prismatic layers at the wall  &  5  &  9    &  12    \\
	Time-step size (in seconds) &  $2 \times 10^{-4}$   &  $1 \times 10^{-4}$    &  $5 \times 10^{-5}$   \\
	\hline
\end{tabular}
\end{center}
\caption{Simulation details}
\label{tab:mesh-convergence-AAA1}
\end{table}

We observe negligible difference in $PSWSS$ amongst the results obtained for the meshes. Therefore, we choose the medium mesh for further analysis.

% Cite papers throughout supporting this
\subsection{Effect of blood constitutive models: Carreau-Yasuda versus Newtonian}
\subsubsection{Velocity contours}
\begin{figure}[h!]
\centering
\includegraphics[width=0.8\linewidth]{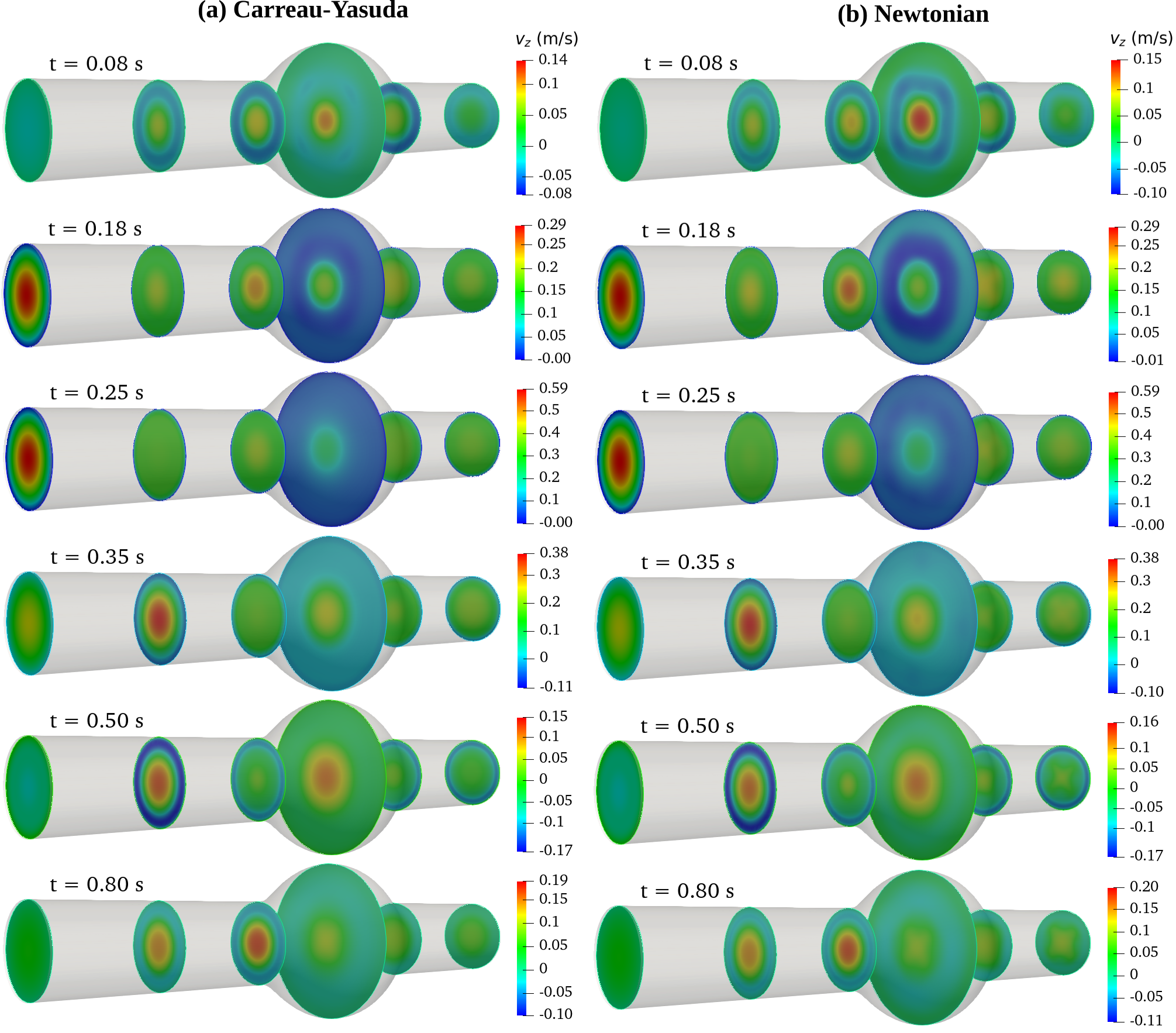}
\caption{Velocity contours at different stations: (a) Carreau-Yasuda and (b) Newtonian models}
\label{fig:rigid-aaa1-velocity-stations}
\end{figure}

The axial velocity ($v_z$) contours plotted at different stations along AAA1 lumen (Fig. \ref{fig:rigid-aaa1-velocity-stations}) for the time instants marked in Fig. \ref{fig:velocity-pressure-waveforms}a suggest that $v_z$ at the central portion of each station is higher for the Newtonian model throughout the cardiac cycle. This is consistent with the fact that dynamic viscosity $\mu^f$ is higher in the regions facing lower shear rate, which typically occurs in the core of the flow, and thus decelerating the flow in those areas.

\subsubsection{Vortex dynamics} \label{subsec:vortex-dynamics-constitutive-models}
%  \begin{figure}[h!]
%  	\centering
%  	\includegraphics[width=\linewidth]{AAA1_lambda2_Carreau}
%  	\caption{}
%  	\label{fig:aaa1-lambda2-carreau}
%  \end{figure}

\begin{figure}[h!]
\centering
\includegraphics[width=0.75\linewidth]{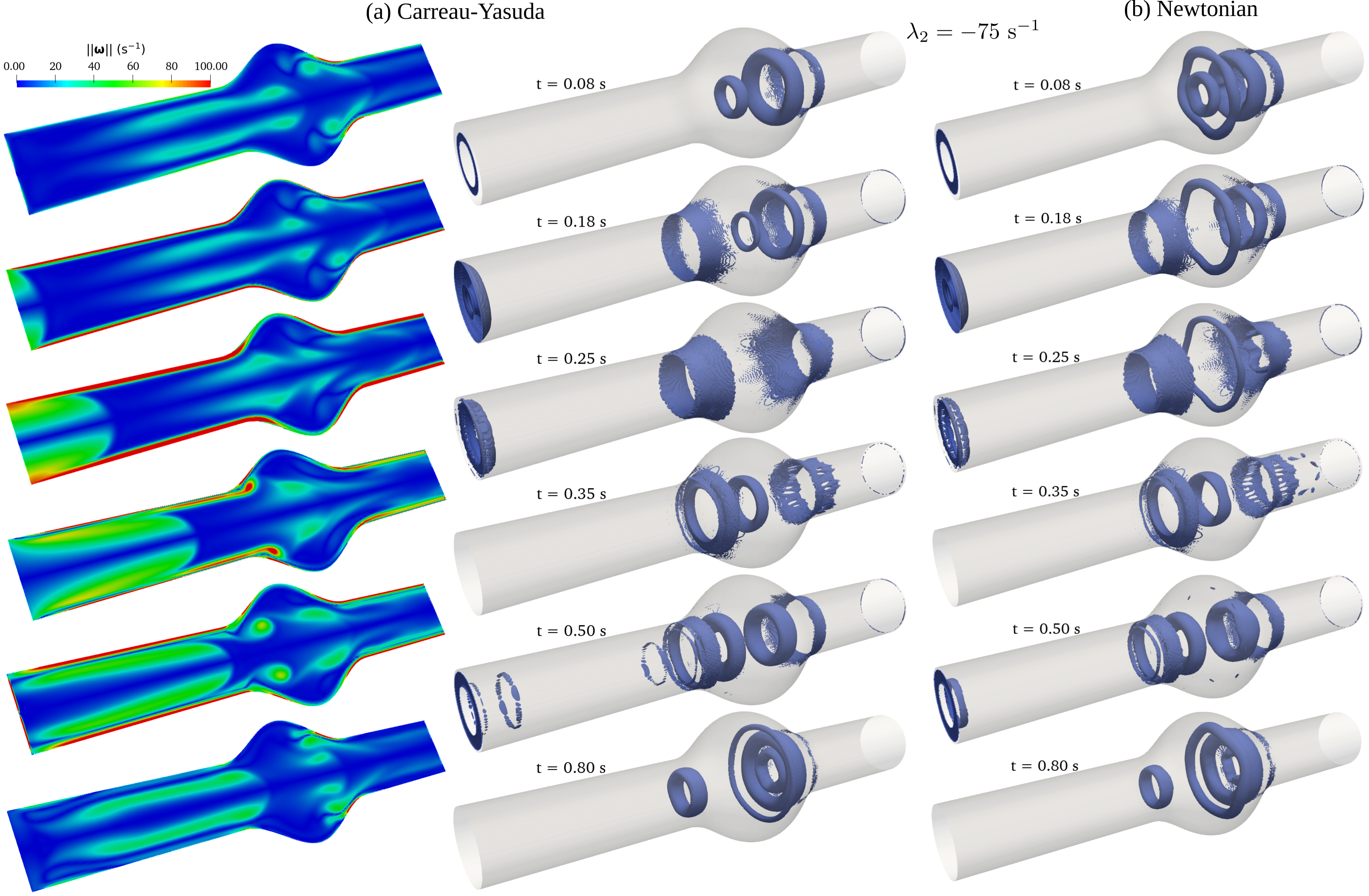}
\caption{Vortex dynamics: (a) Vorticity contours and $\lambda_2 = -75$ s$^{-2}$ iso-surface for Carreau-Yasuda model, (b) $\lambda_2 = -75$ s$^{-2}$ iso-surface for Newtonian model.}
\label{fig:aaa1-lambda2-constitutive-models}
\end{figure}

A presentation of vortex dynamics in Fig. \ref{fig:aaa1-lambda2-constitutive-models}a displays strong correlation between the regions of high values of vorticity $\vb*{\omega} ~( = \curl{\vb*{v}}^f)$ and ring vortices obtained using $\lambda_2$-criterion for a fixed $\lambda_2 = -75$ s$^{-2}$ for Carreau-Yasuda model. As has been discussed in Finol et al. \cite{Finol2003,Finol2003a}, the growth, decay and transport of vortices in the flow can be divided into the following flow phases alongwith the corresponding time intervals in the cardiac cycle
\begin{itemize}
\item \textit{Systolic acceleration} ($t = 0.08$ s to $t = 0.25$ s)  involves downstream transport of the residual vortices, left from the previous cardiac cycle, out of the aneurysms, resulting in a completely attached flow pattern. 

\item \textit{Systolic deceleration} ($t = 0.25$ s to $t = 0.50$ s) is characterized by flow separation at the proximal neck of the aneurysm, the growth of a single-vortex near the midsection which then moves downstream. 

\item \textit{Early diastole} ($t = 0.50$ s to $t = 0.85$ s) involves gradual weakening of the vortex shed at the end of systole as it translates downstream and the growth of the translating single-vortex.

\item \textit{Late diastole}  ($t = 0.85$ s to $t = 0.08$ s) is the interval where the most significant
flow disturbance takes place, mainly due to the low-velocity recirculation region inside the aneurysm towards its distal end. Also, the downstream vortex translation leads to enclosing of
the forward flow stream by the vortex.
\end{itemize}
The vortex dynamics for Newtonian case differs slightly from that of the Carreau-Yasuda in that there is an occurrence of a weak wavy-like ring vortex of large diameter in all but systolic deceleration phase as seen in Fig. \ref{fig:aaa1-lambda2-constitutive-models}a and \ref{fig:aaa1-lambda2-constitutive-models}b.  

\subsubsection{WSS and derived quantities}
\begin{figure}[h!]%
\centering
\subfloat[\centering ]{{\includegraphics[width=0.49\textwidth]{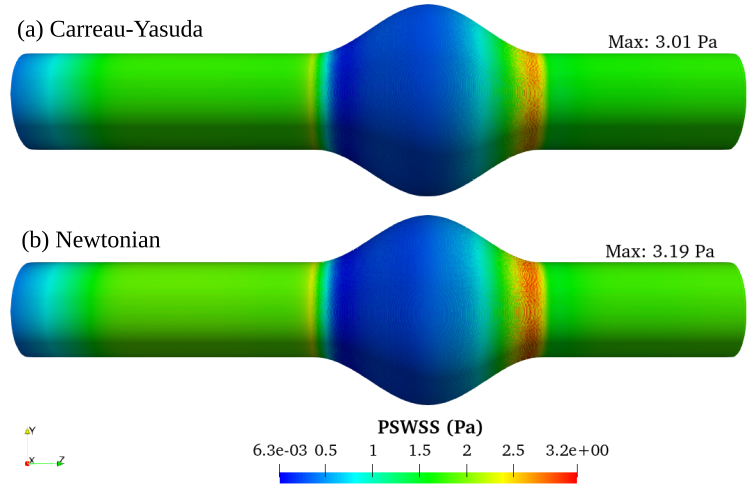} } \label{fig:rigid-aaa1-constitutive-models-pswss}}%
\qquad
\subfloat[\centering ]{{\includegraphics[width=0.49\textwidth]{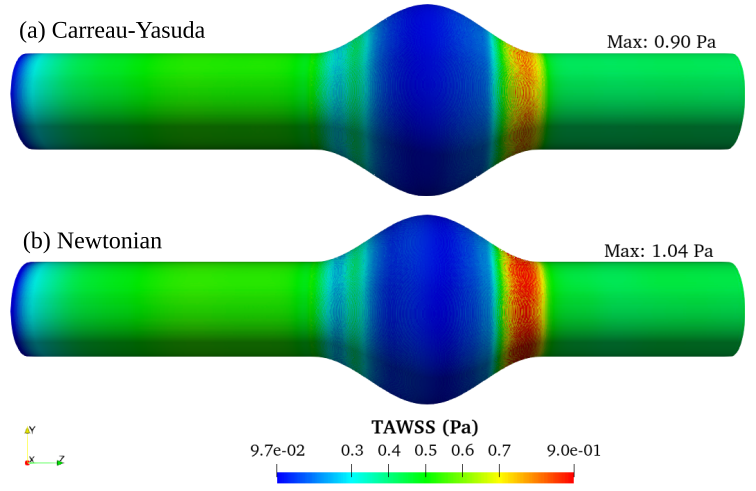} } \label{fig:rigid-aaa1-constitutive-models-tawss}}%
\caption{(left) $PSWSS$ and (right) $TAWSS$ distributions for (a) Carreau-Yasuda and (b) Newtonian models for AAA1}%
\label{fig:rigid-aaa1-constitutive-models-pswss-tawss}%
\end{figure}

It is clear from Fig. \ref{fig:rigid-aaa1-constitutive-models-pswss} that $PSWSS$ values for Newtonian case are higher than the Carreau-Yasuda case throughout the domain and the difference is prominently noticeable in the portion towards the distal end of the aneurysm sac; the maximum $PSWSS$ values reaching 3.19 Pa and  3.01 Pa respectively. This is so because the dynamic viscosity is lower in the wall region for the Carreau-Yasuda case which produces lower wall shear stress.

Similar to $PSWSS$, Newtonian model overpredicts $TAWSS$ compared to Carreau-Yasuda model  (see Fig. \ref{fig:rigid-aaa1-constitutive-models-tawss}) with higher variations in the distal region of the aneurysm. While the maximum $TAWSS$ in the case of Carreau-Yasuda model reaches  0.9 Pa, the corresponding value for Newtonian model is 1.04 Pa; an increase of $16 \%$.

\begin{figure}[h!]
\centering
\includegraphics[width=0.5\linewidth]{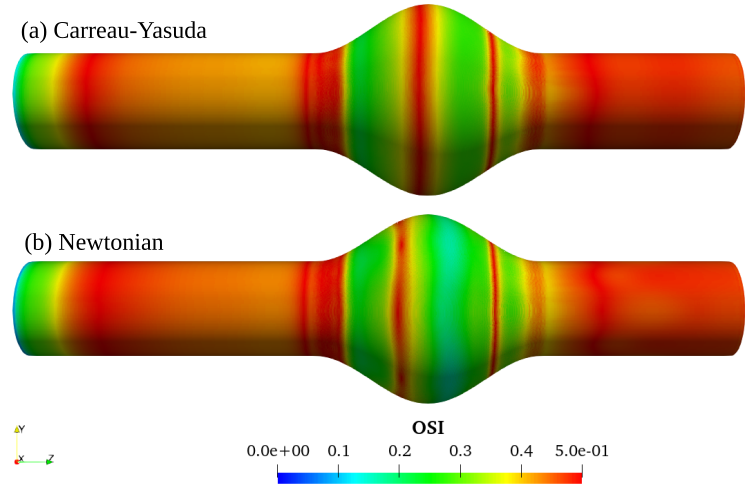}
\caption{$OSI$ distribution: (a) Carreau-Yasuda and (b) Newtonian models}
\label{fig:rigid-aaa1-constitutive-models-osi}
\end{figure}

We observe from Fig. \ref{fig:rigid-aaa1-constitutive-models-osi} that blood flow undergoes higher fluctuations in the aneurysm sac for Carreau-Yasuda case. It is also noticed that while high $OSI$ values are found in the central portion of the aneurysm sac, the high $OSI$ region forming a narrow band shifts proximally in the Newtonian case. 

%\cleardoublepage
\subsection{Effect of DHr: AAA1 versus AAA2}
We study how the flow dynamics is influenced by the shape index DHr - AAA1 with a higher DHr and AAA2 with a lower DHr. For both cases, simulations have been performed using Carreau-Yasuda blood model.

\subsubsection{Velocity contours}
\begin{figure}[h!]
\centering
\includegraphics[width=0.6\linewidth]{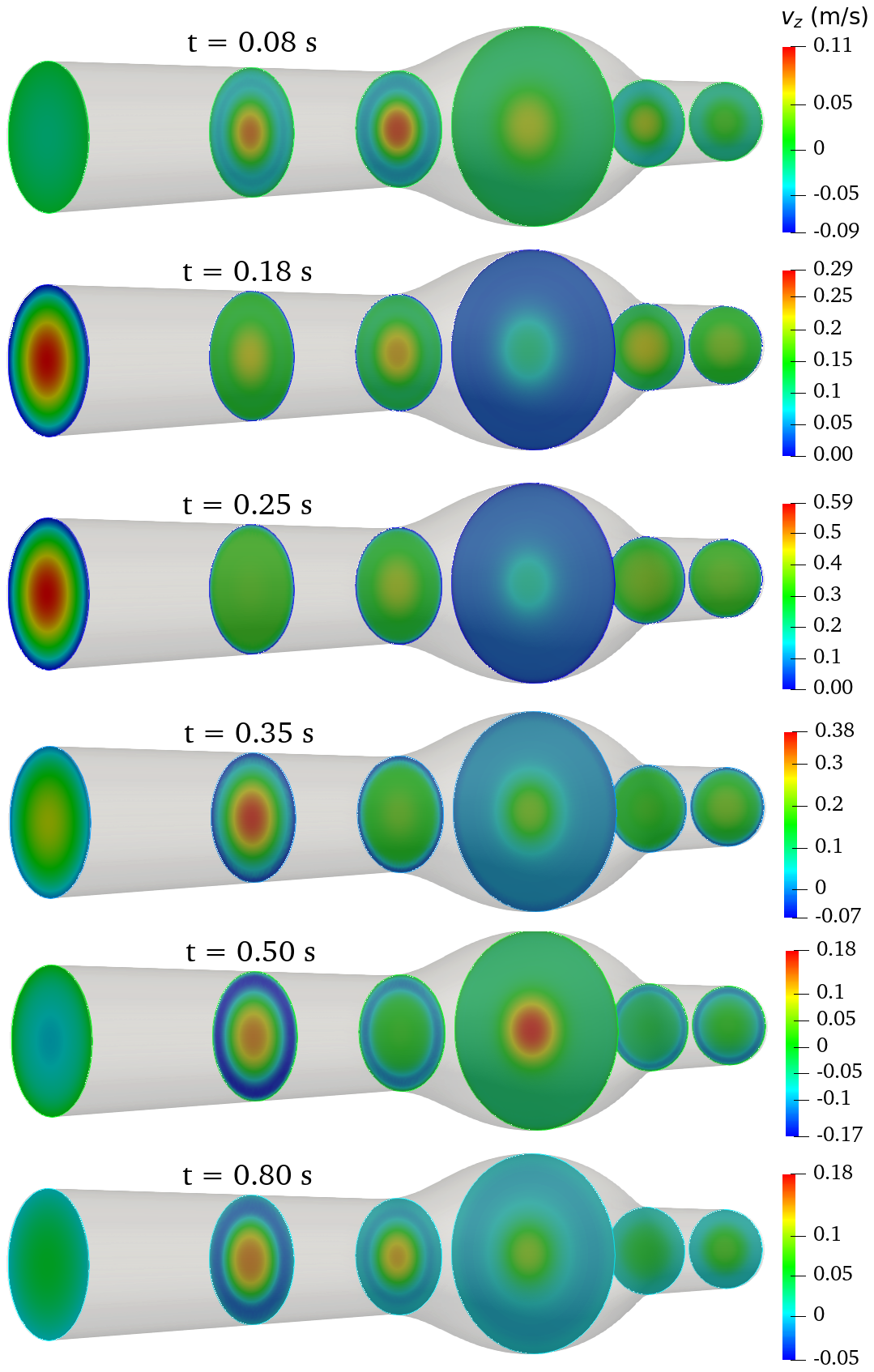}
\caption{Velocity contours at different stations of AAA2}
\label{fig:rigid-aaa2-velocity-contours}
\end{figure}

Fig. \ref{fig:rigid-aaa2-velocity-contours} shows the axial velocity ($v_z$) contours plotted at different stations along AAA2 lumen for various time instants. The velocity at the maximum bulge location is relatively lower than other stations for all time instants except t = 0.5 s. %We find the same trend in the case of AAA1 too (Fig. \ref{fig:rigid-aaa1-velocity-constitutive-models}).

The flow dynamics inside AAA1 forms larger recirculation zones and flow separation than that of AAA2. The difference in the extent of recirculation regions are conspicuous at t = 0.08, 0.18, 0.8 s. In such regions, thrombus finds a fertile environment to settle down and gradually increase in size. Therefore, there is an increased chance of ILT deposition in AAA1 compared to AAA2.

\begin{figure}[h!]
\centering
\includegraphics[width=0.6\linewidth]{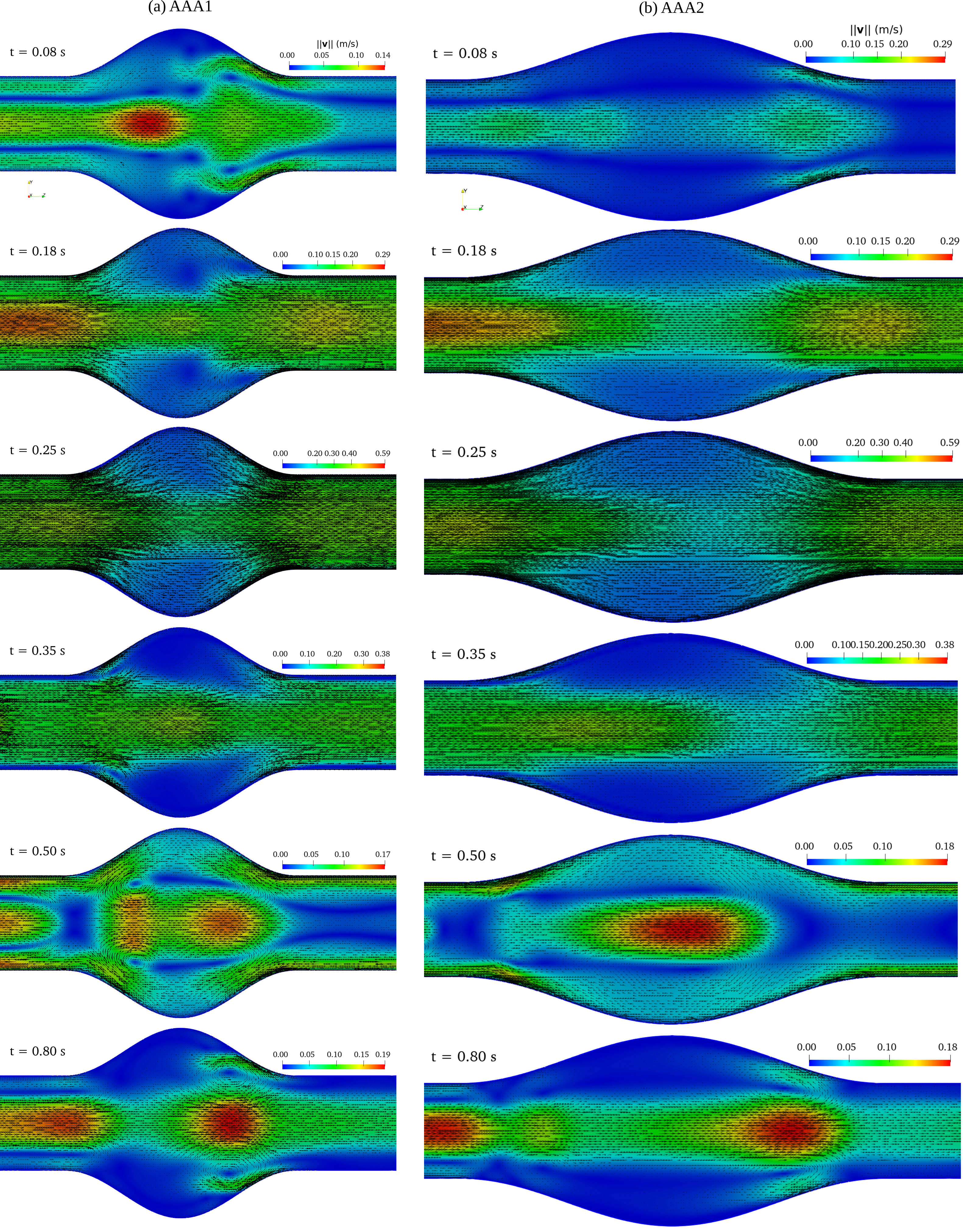}
\caption{Velocity vector plot in the longitudinal plane of the aneurysm sac: (a) AAA1 (b) AAA2}
\label{fig:rigid-different-dhr-velocity}
\end{figure}

\subsubsection{Pressure contours}

\begin{figure}[h!]
\centering
\includegraphics[width=1.0\linewidth]{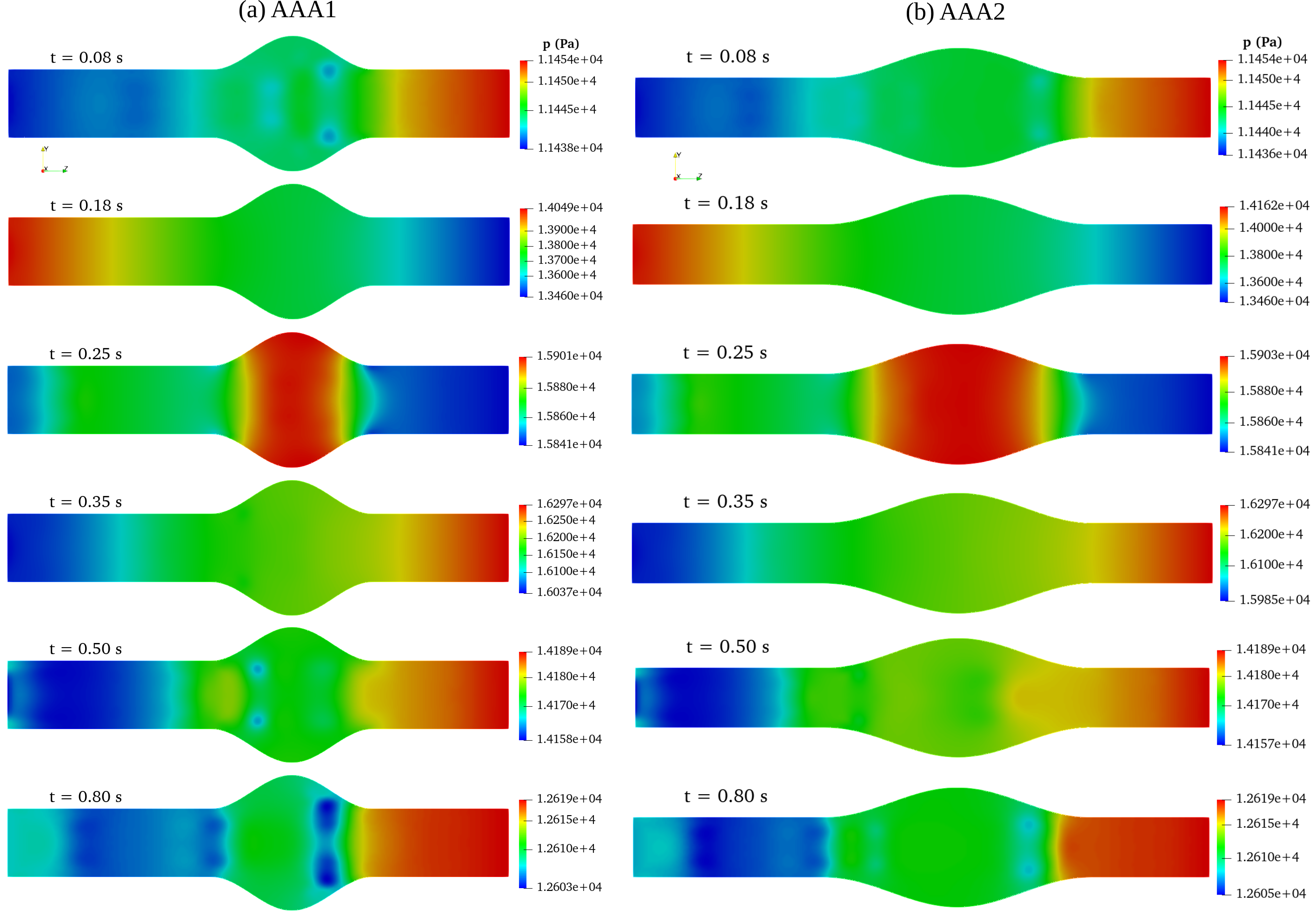}
\caption{Pressure distribution in the longitudinal plane for (a) AAA1 and (b) AAA2}
\label{fig:rigid-different-dhr-pressure}
\end{figure}

We observe from Fig. \ref{fig:rigid-different-dhr-pressure} that the temporal  acceleration and deceleration of the flow has a major influence on the pressure distribution throughout the cardiac cycle in both AAAs. The flow sees strong favourable pressure gradient in the initial systolic phase, that is, between $t = 0.1$ s to $t = 0.25$ s. For the remaining interval of the cardiac cycle, the flow suffers from adverse pressure gradient and is strongest around $t = 0.35$ s. 

The pressure distribution for AAA1 and AAA2 appears to remain fairly similar except for symmetric pockets of lower pressure found in the aneurysm sac of AAA1 at $t = 0.08$ s, $t = 0.5$ s and $t = 0.8$ s which are absent in AAA2. These pockets can be correlated with ring vortices formed in AAA1. %Our results are consistent with the findings of Finol et al. \cite{Finol2001}.  

\subsubsection{Vortex dynamics}
\begin{figure}[h!]
\centering
\includegraphics[width=0.75\linewidth]{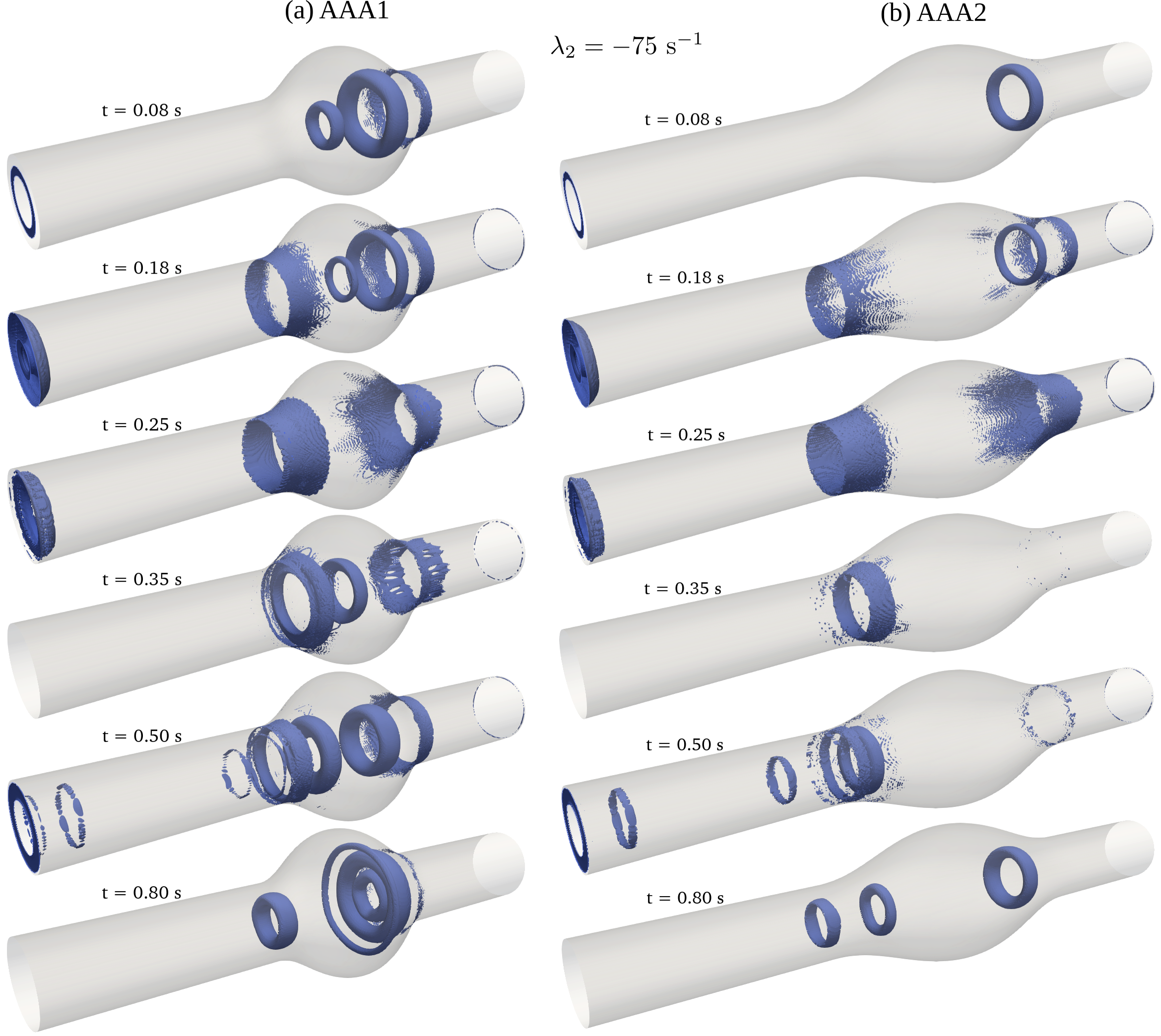}
\caption{Vortex dynamics: $\lambda_2 = -75$ s$^{-2}$ iso-surface contours for (a) AAA1 (b) AAA2.}
\label{fig:aaa-lambda2-different-dhr}
\end{figure}

The vortex dynamics for AAA2 can be categorised as discussed earlier in Section \ref{subsec:vortex-dynamics-constitutive-models} and display similar behaviour. For $\lambda_2 = -75$ s$^{-2}$ iso-surface shown in Fig. \ref{fig:aaa-lambda2-different-dhr}, we note the following distinctions pertaining to the vortex dynamics in the flow phases of AAA1 and AAA2
\begin{itemize}
\item \textit{Systolic acceleration} ($t = 0.08$ s to $t = 0.25$ s): A weak vortex is observed at the distal end of AAA2 unlike two vortices found at the mid-way and distal end of AAA1 sac. In both, the vortices continue to decay till they vanish at peak systole.

\item \textit{Systolic deceleration} ($t = 0.25$ s to $t = 0.50$ s): The growth of two vortices formed at the proximal end and mid-way of AAA1 sac takes place; both translates downstream. A weak vortex forms in the proximal end of AAA2 sac which moves downstream too. 

\item \textit{Early diastole} ($t = 0.50$ s to $t = 0.85$ s): While significant flow disturbance takes place in AAA1 causing growth, decay and translation of multiple vortices, the situation in AAA2 is relatively quiet.

The gradual weakening of the vortex shed at the end of systole as it translates downstream and the growth of the translating single-vortex.

\item \textit{Late diastole}  ($t = 0.85$ s to $t = 0.08$ s):  The downstream vortex translation leads to enclosing of the forward flow stream by the vortex in AAA1. However, decay of two vortices at the proximal end of AAA2 takes place accompanied by a gradual narrowing of the distal vortex. 
\end{itemize}

In a nutshell, AAA1 has richer and more proactive vortex dynamics than AAA2.  %As DHr increases, we find that vortex formation happens more actively 

\subsubsection{WSS and derived quantities}

%Finol et al. \cite{Finol2003}

%Boyd et al. \cite{Boyd2016} This CFD study was the ﬁrst to model aortic blood ﬂow in realistic geometries of RAAAs. In contradiction to our original hypothesis, rupture location coincided with regions of predicted low WSS and ﬂow recirculation, not with regions of high pressure and WS as predicted by FE analysis. The reason for this discrepancy is not yet known. Future studies examining the interplay between alterations in WSS in pulsatile ﬂow and its effect on vascular endothelial remodeling will lead to a better understanding of AAA development and growth and may ultimately provide better prediction of AAA rupture   potential.

\begin{figure}[h!]%
\centering
\subfloat{{\includegraphics[width=0.49\textwidth]{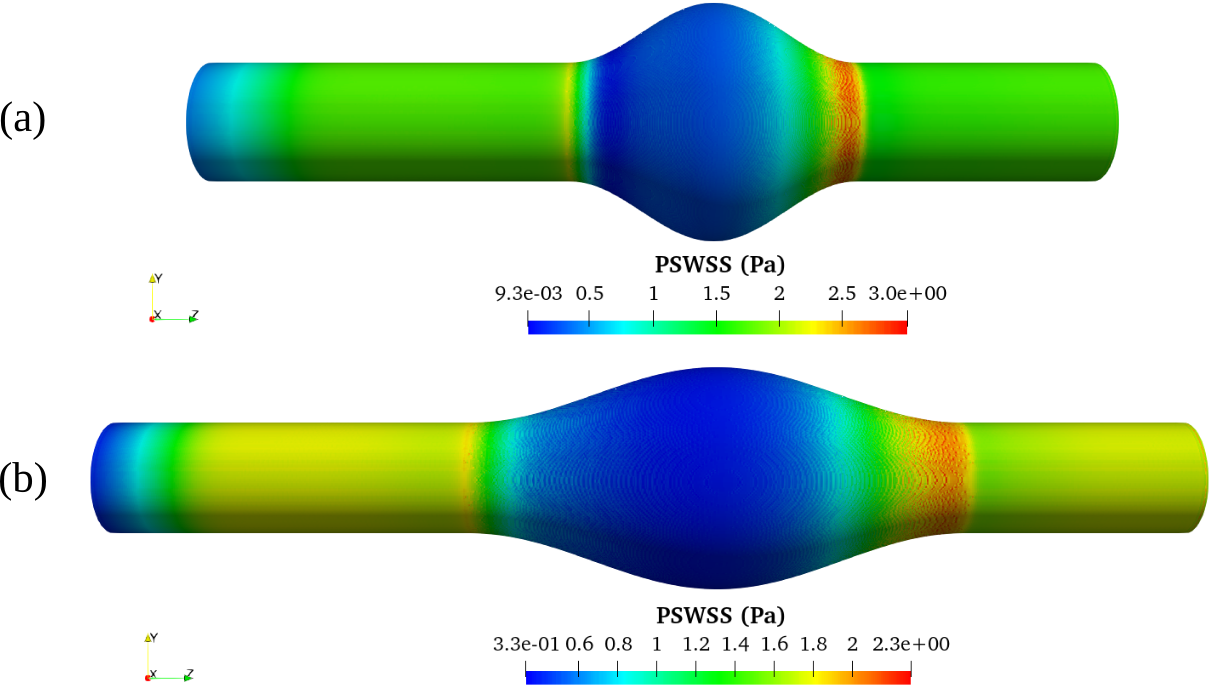} } \label{fig:rigid-different-dhr-pswss}}%
\qquad
\subfloat{{\includegraphics[width=0.49\textwidth]{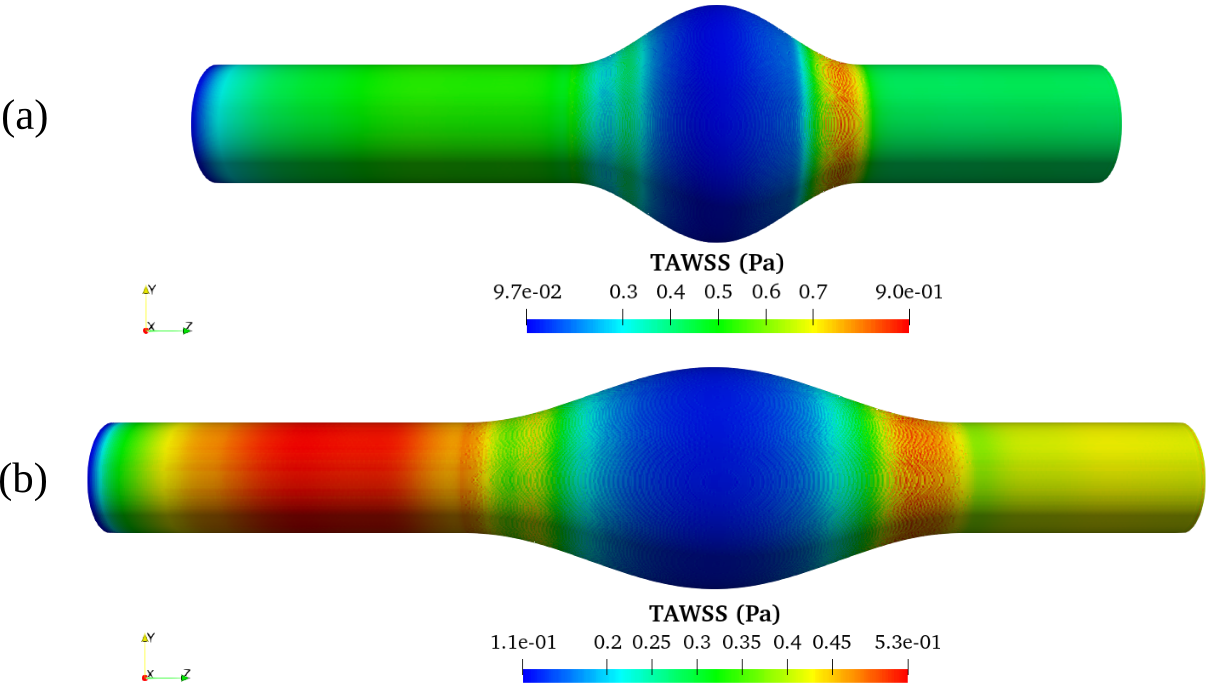} } \label{fig:rigid-different-dhr-tawss}}%
\caption{ $PSWSS$ and $TAWSS$ distributions for (a) AAA1 and (b) AAA2}%
\label{fig:rigid-different-dhr-pswss-tawss}%
\end{figure}

$PSWSS$ varies from 0.0093 to 3 Pa in the case of AAA1 and 0.33 to 2.3 Pa for AAA2 (Fig. \ref{fig:rigid-different-dhr-pswss-tawss}). In both the cases, the maximum $PSWSS$ occurs at the corresponding distals ends. %Likewise, $meanWSS$ is maximum at the distal end of AAA1 and at the proximal end of AAA2, though the maximum $meanWSS$ is higher in the case of AAA1 (0.29 Pa) than AAA2 (0.23 Pa). 

% meanWSS
%\begin{figure}[h!]
%	\centering
%	\includegraphics[width=0.65\linewidth]{rigid_different_DHr_meanWSS}
%	\caption{meanWSS distribution for (a) AAA1 and (b) AAA2}
%	\label{fig:rigid-different-dhr-meanwss}
%\end{figure}

It can be observed Fig. \ref{fig:rigid-different-dhr-pswss-tawss} that $TAWSS$ varies from 0.097 to 0.9 Pa in AAA1, AAA2 in the range 0.11 to 0.53 Pa. AAA1 observes lower values of $TAWSS$ than AAA2 in the region left to its maximum bulge. In either cases, we find that coexisting high and low $TAWSS$ regions persists towards the proximal end (and even at the distal end) creating a fertile ground for the thrombus formation inside the AAAs. This is more apparent in AAA1 case. %Initiation of thrombus deposition occurs at high $TAWSS$ \cite{Philip2020}

%\begin{figure}[h!]
%	\centering
%	\includegraphics[width=0.65\linewidth]{rigid_different_DHr_PSWSS}
%	\caption{PSWSS distribution for (a) AAA1 and (b) AAA2}
%	\label{fig:rigid-different-dhr-pswss}
%\end{figure}
%
%\begin{figure}[h!]
%	\centering
%	\includegraphics[width=0.65\linewidth]{rigid_different_DHr_TAWSS}
%	\caption{$TAWSS$ distribution for (a) AAA1 and (b) AAA2}
%	\label{fig:rigid-different-dhr-tawss}
%\end{figure} 

\begin{figure}[h!]%
\centering
\subfloat{{\includegraphics[width=0.49\textwidth]{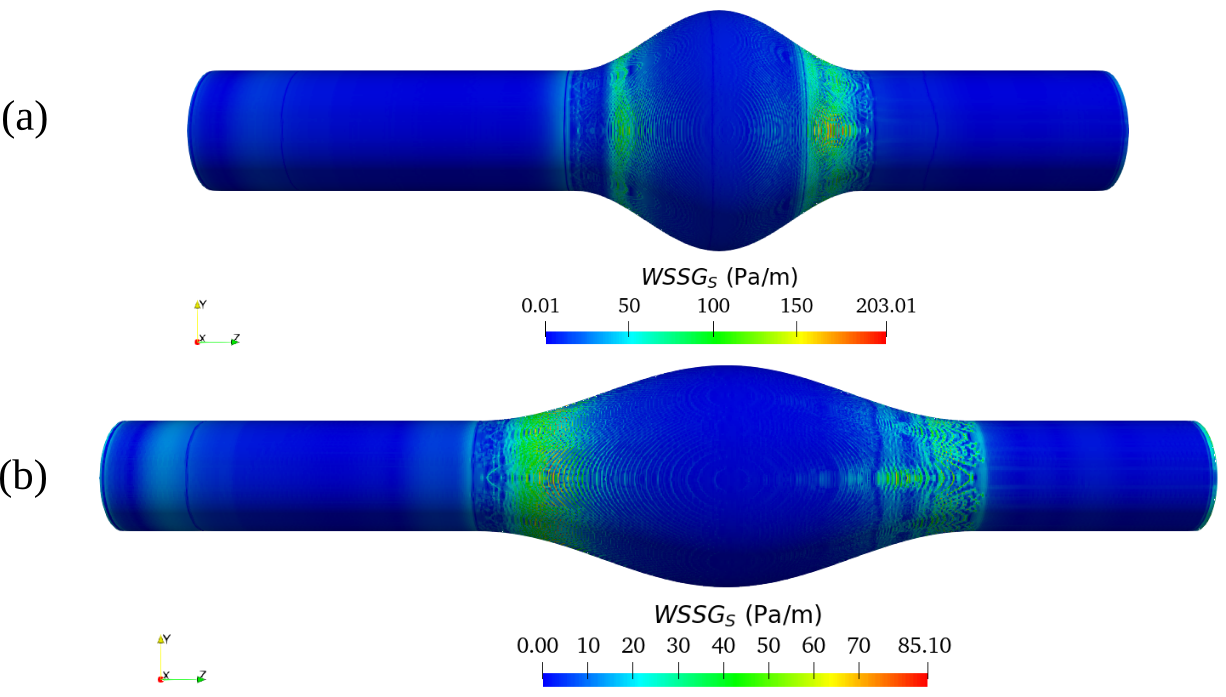} } \label{fig:rigid-different-dhr-wssgs}}%
\qquad
\subfloat{{\includegraphics[width=0.49\textwidth]{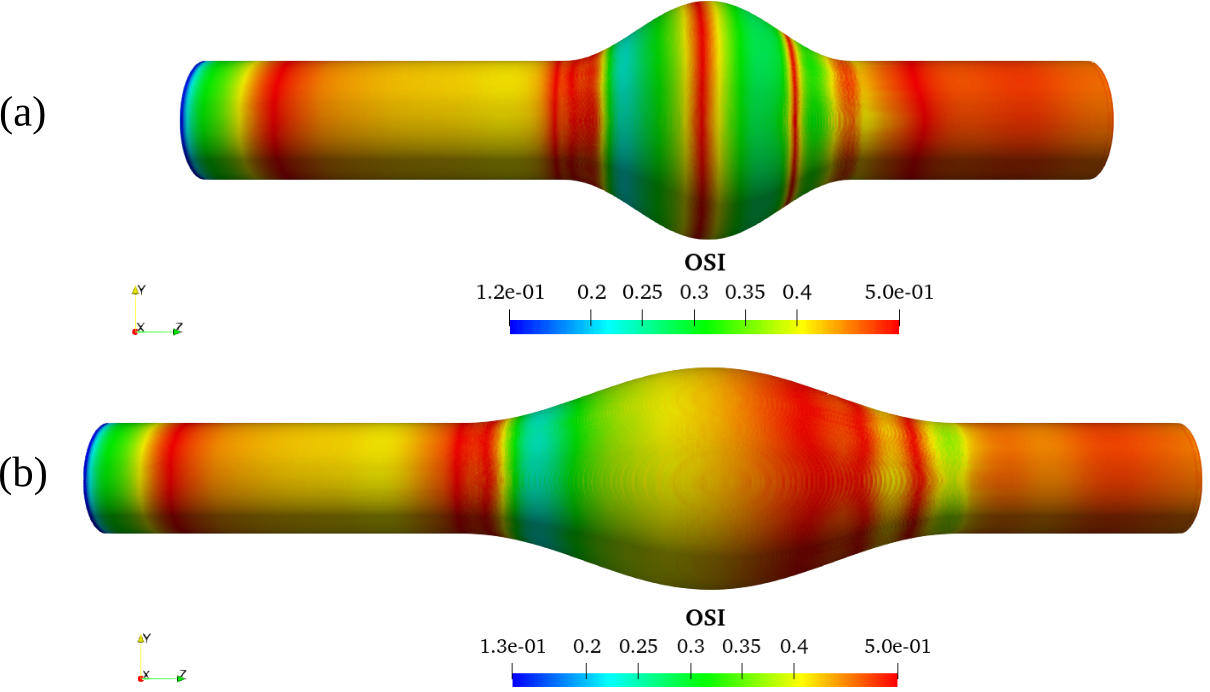} } \label{fig:rigid-different-dhr-osi}}%
\caption{ $WSSG_S$ and  $OSI$ distributions for (a) AAA1 and (b) AAA2}%
\label{fig:rigid-different-dhr-wssgs-osi}%
\end{figure}

% WSSGS
AAA1 suffers from higher maximum $WSSG_S$ (203.1 Pa/m) than AAA2 (85.10 Pa/m); an increase by a factor of 2 than AAA2 (Fig. \ref{fig:rigid-different-dhr-wssgs-osi}). In either case, the maximum happens to be located at the proximal and distal ends. The high $OSI$ region in the central portion of AAA1, flanked by lower $OSI$ regions on either sides (Fig. \ref{fig:rigid-different-dhr-wssgs-osi}), suggests that the flow velocity undergoes severe fluctuations for the  maximum bulge location. However, the situation is quite different for AAA2 in which $OSI$ progressively increases from its proximal end to its distal end.

%\begin{figure}[h!]
%	\centering
%	\includegraphics[width=0.65\linewidth]{rigid_different_DHr_WSSGS}
%	\caption{$WSSG_S$ distribution for (a) AAA1 and (b) AAA2}
%	\label{fig:rigid-different-dhr-wssgs}
%\end{figure}
%
%\begin{figure}[h!]
%	\centering
%	\includegraphics[width=0.65\linewidth]{rigid_different_DHr_OSI}
%	\caption{$OSI$ distribution for (a) AAA1 and (b) AAA2}
%	\label{fig:rigid-different-dhr-osi}
%\end{figure}

\section{Computational solid stress simulations of arterial wall mechanics}
To assess stress in the arterial wall, we use von Mises stress and first principal stress. The state of stress at every material point in the arterial wall is defined in terms of Cauchy stress tensor $\vb*{\sigma}^s$ which is symmetric and consists of 6 components, namely, $\sigma_{ij}^s$ ($i,j = x, y, z$). We quantify the stress state at every point in terms of von Mises stress and maximum principal stress $\sigma_1$.  The von Mises stress $\sigma_{eq}$ is defined as
\begin{align*}
	\sigma_{eq} = \frac{1}{\sqrt{2}}\sqrt{(\sigma_{xx} - \sigma_{yy})^2 + (\sigma_{yy} - \sigma_{zz})^2 + (\sigma_{xx} - \sigma_{zz})^2 + 6 (\sigma_{xy}^2 + \sigma_{yz}^2 + \sigma_{xz}^2)}
\end{align*}
while the maximum principal stress $\sigma_1$ is the maximum eigenvalue of $\vb*{\sigma}^s$.
%\[
%\vb*{\sigma}^s =
%\begin{bmatrix}
%\sigma_{xx} & \sigma_{xy} & \sigma_{xz} \\
%\sigma_{xy} & \sigma_{yy} & \sigma_{yz} \\
%\sigma_{xz} & \sigma_{yz} & \sigma_{zz}
%\end{bmatrix}
%\]

We perform computational solid stress (CSS$_T$) simulations for both AAA1 and AAA2 walls. In CSS$_T$, a time-varying uniform pressure loading corresponding to the cardiac cycle (see Fig. \ref{fig:velocity-pressure-waveforms}) is applied on the endoluminal surface of the aneurysmal walls. We study the influence of solid material models and $DHr$ on the wall mechanics.
%  \begin{figure}[h!]
%  	\centering
%  \includegraphics[width=\linewidth]{solidWall_AAA1_coarse_Poisson_ratio}
%  	\caption{Effect of Poisson's ratio on von Mises stress $\sigma_{eq}$ (left) and displacement $D$ (right) for (a) $\nu = 0.3$ (b) $\nu = 0.4$ (c) $\nu = 0.45$. Gray colour in the displacement indicates undeformed configuration}
%  	\label{fig:solidwall-aaa1-coarse-poisson-ratio}
%  \end{figure}

\begin{figure}[h!]%
\centering
\subfloat[\centering ]{{\includegraphics[width=0.45\textwidth]{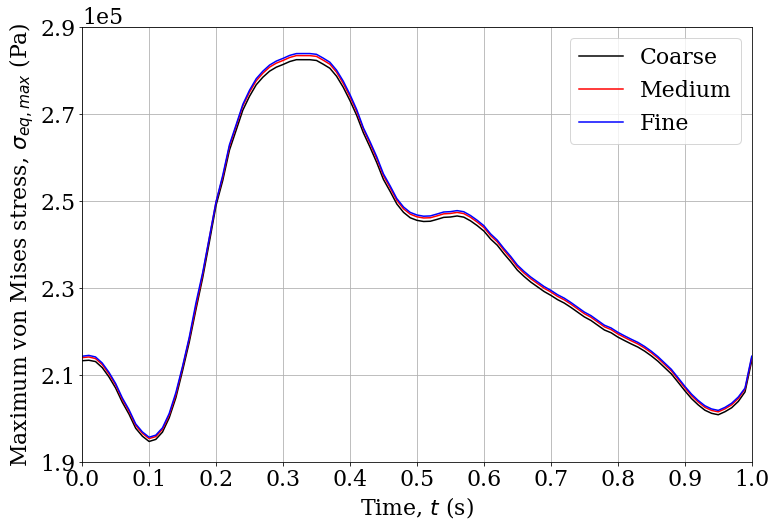} } \label{fig:mesh-sensitivity-solid-wall-aaa1-linear-elastic}}%
\qquad
\subfloat[\centering ]{{\includegraphics[width=0.45\textwidth]{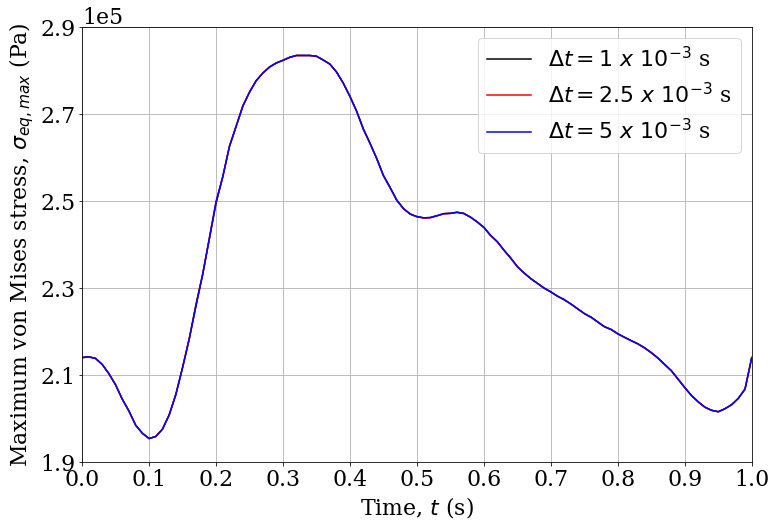} } \label{fig:time-sensitivity-solid-wall-aaa1-linear-elastic}}%
\caption{AAA1 wall modeled as a linear elastic material: (a) Mesh and (b) Temporal convergence}%
\label{fig:mesh-and-time-sensitivity-solid-wall-aaa1-linear-elastic}%
\end{figure}

%\subsection{Convergence study}
%\paragraph{Mesh convergence}
For spatial convergence studies, we consider three different mesh resolutions for AAA1 wall (see Table \ref{tab:spatial-convergence-CSST}) using time-step size of $5\times 10^{-3}$ s.% Although the governing equations for solid dynamics use total Lagrangian formulation, we still allow the simulation to run for two cardiac cycles to achieve solution periodicity.
\begin{table}[h!]
\begin{center}
\begin{tabular}{|c|c|c|}
	\hline
	& Number of cells in radial direction & Number of cells  \\
	\hline
	Coarse & 3   & 31,500  \\
	\hline
	Medium  &  5   & 73,800  \\
	\hline
	Fine &  8 & 1,63,700 \\
	\hline
\end{tabular}
\end{center}
\caption{Mesh information}
\label{tab:spatial-convergence-CSST}
\end{table}
It can be inferred from Fig. \ref{fig:mesh-sensitivity-solid-wall-aaa1-linear-elastic} that there is negligible difference between the maximum von Mises stress corresponding to the medium and fine meshes.

%\paragraph{Temporal convergence}
For the medium mesh, we consider three different time-step sizes for time integration, namely, $5\times 10^{-3}$ s, $2.5 \times 10^{-3}$ s and $1.0 \times 10^{-3}$ s. We infer from Fig. \ref{fig:time-sensitivity-solid-wall-aaa1-linear-elastic} that the influence of time-step size on the solution, here maximum von Mises stress in the wall, is negligible.  

On the basis of spatial and temporal convergence studies, we choose the medium mesh for further simulations on AAA1 wall using a time-step of $2.5 \times 10^{-3}$ s. The mesh settings utilised to discretise AAA1 wall is used for AAA2 wall as well.

\begin{figure}[h!]%
\centering
\subfloat[\centering ]{{\includegraphics[width=0.49\textwidth]{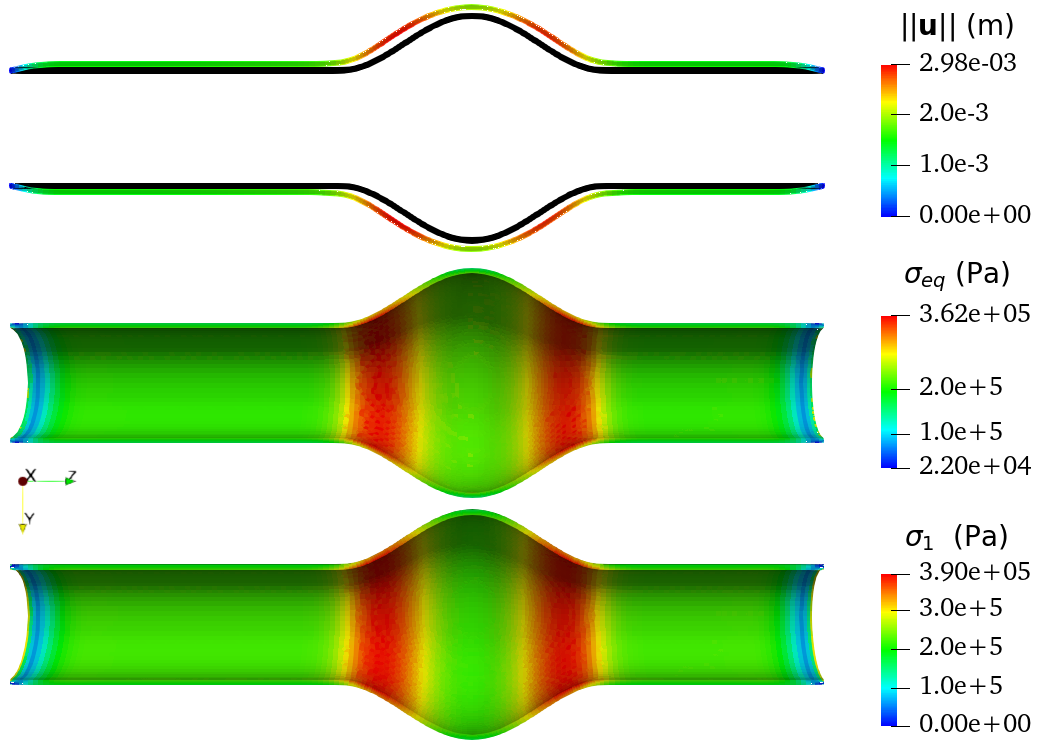} } \label{fig:solid-wall-aaa1-raghavan-vorp-medium}}%
\qquad
\subfloat[\centering ]{{\includegraphics[width=0.49\textwidth]{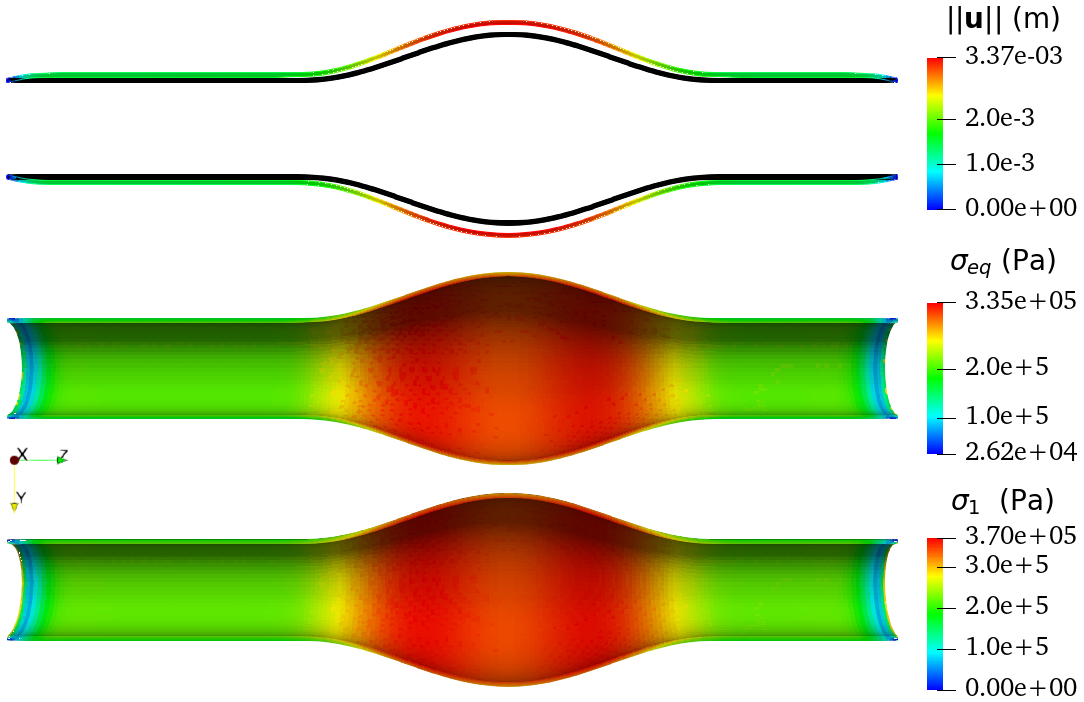} } \label{fig:solid-wall-aaa2-raghavan-vorp-medium}}%
\caption{Total displacement (top), von Mises stress (middle), first principal stress (bottom) for (a) AAA1 and (b) AAA2 walls modeled as Raghavan-Vorp elastic material at $t = 0.32$ s under peak uniform pressure loading}%
\label{fig:solid-wall-aaa1-aaa2-raghavan-vorp-medium}%
\end{figure}

%\subsection{Influence of solid material models and DHr}
Fig. \ref{fig:solid-wall-aaa1-raghavan-vorp-medium} shows the total displacement $\norm{\vb*{u}}$, von Mises stress $\sigma_{eq}$ and the first principal stress $\sigma_1$ for AAA1 wall modeled as Raghavan-Vorp elastic model at $t=0.32$ s, that is the peak time of the uniform pressure loading. The displacement is highest ($2.98$ mm) in the midway of the initiation of curvature and the maximum bulge location. A similar observation is made regarding $\sigma_{eq}$ and $\sigma_1$ whose maximum values are $3.62 \times 10^5$ Pa and $3.90 \times 10^5$ Pa respectively. The region flanking the central portion of the bulge sees reduced values for all these variables. The maximum displacement is about 3.37 mm for AAA2 wall (see Fig. \ref{fig:solid-wall-aaa2-raghavan-vorp-medium}), higher than the corresponding value for AAA1 wall. Likewise, the peak values for $\sigma_{eq}$ and $\sigma_1$ are $3.35 \times 10^5$ Pa and $3.70 \times 10^5$ Pa respectively, lower than the corresponding values for AAA1 wall. Since the peak wall stress is higher in the case of AAA1, this keeps it at a higher risk of rupture than AAA2.

%Our $\sigma_1$ results lie within the range as obtained by Perez et al. \cite{Perez2016}  who performed a steady state FEM analysis conducted on 190 AAA geometries with the endoluminal surface of the AAAs subjected to uniform peak pressure of the cardiac cycle and obtained an average maximum $\sigma_1$ to be $5.6 \times 10^5$ Pa ($2.2 \times 10^5$ - $1.18 \times 10^6$ Pa). 

%\begin{figure}[h!]
%	\centering
%	\includegraphics[width=\linewidth]{solidWall_AAA1_raghavanVorp_medium}
%	\caption{Total displacement (top), von Mises stress (middle), first principal stress (bottom) for AAA1 wall modeled as Raghavan-Vorp elastic material at $t = 0.32$ s under peak uniform pressure loading}
%	\label{fig:solid-wall-aaa1-raghavan-vorp-medium}
%\end{figure}
%
%\begin{figure}[h!]
%	\centering
%	\includegraphics[width=\linewidth]{solidWall_AAA2_raghavanVorp_medium}
%	\caption{Total displacement (top), von Mises stress (middle), first principal stress (bottom) for AAA2 wall modeled as Raghavan-Vorp elastic material at $t = 0.32$ s under peak uniform pressure loading}
%	\label{fig:solid-wall-aaa2-raghavan-vorp-medium}
%\end{figure}

Since the trend of variation of $\norm{\vb*{u}}$, $\sigma_{eq}$ and $\sigma_1$ differ only quantity-wise for each material model but not quality-wise, we shall plot the peak values of these quantities over a cardiac cycle for both AAA1 and AAA2 cases. 

\subsubsection{Influence of constitutive models for solid}

\begin{figure}[h!]
\centering
\includegraphics[width=0.5\linewidth]{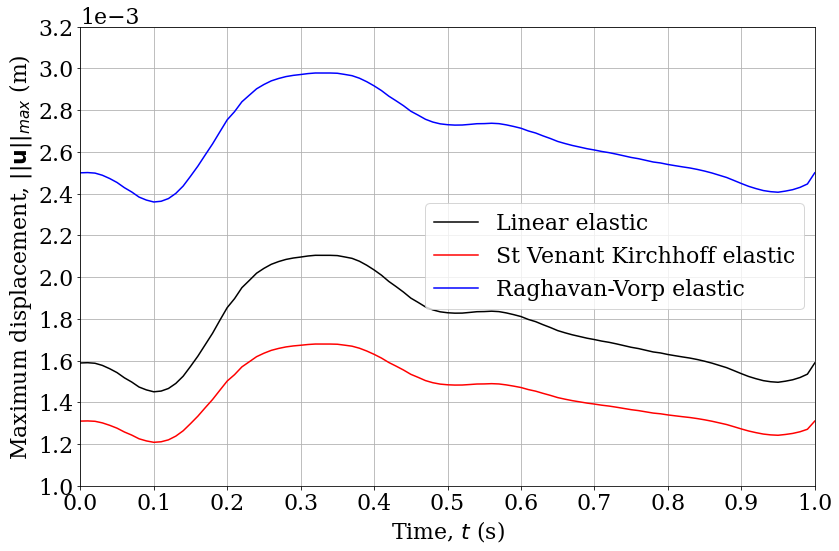}
\caption{Variation of maximum total displacement in AAA1 wall over a cardiac cycle for linear, Saint Venant Kirchhoff and Raghavan-Vorp elastic models}
\label{fig:solidwall-aaa1-max-displacement}
\end{figure}

\begin{figure}[h!]%
\centering
\subfloat[\centering ]{{\includegraphics[width=0.48\textwidth]{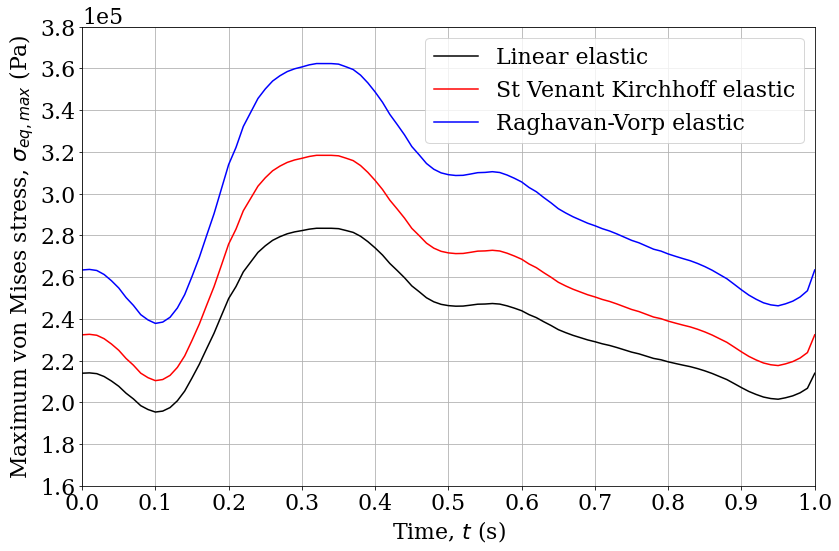} } \label{fig:solidwall-aaa1-max-von-mises-stress}}%
\qquad
\subfloat[\centering ]{{\includegraphics[width=0.48\textwidth]{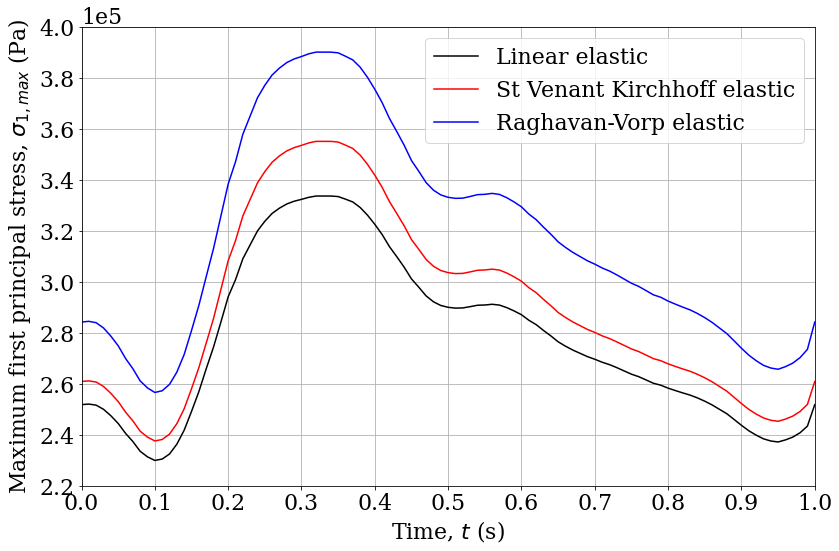} } \label{fig:solidwall-aaa1-max-principal-stress}}%
\caption{Variation of (a) maximum principal stress and (b) maximum von Mises stress in AAA1 wall over a cardiac cycle for linear, Saint Venant Kirchhoff and Raghavan-Vorp elastic models}%
\label{fig:solidwall-aaa1-max-von-mises-principal-stress}%
\end{figure}

The maximum displacement of the AAA1 wall in the  Raghavan-Vorp elastic model ($\approx 3$ mm) is about 1.4 times that of linear elastic ($\approx 2.1$ mm) and approximately 1.8 times that of Saint Venant Kirchhoff elastic models (Fig. \ref{fig:solidwall-aaa1-max-displacement}). Maximum von Mises stress in AAA1 wall for Raghavan-Vorp elastic model ($\approx 3.60 \times 10^5$ Pa) is 12.4 \% higher than the corresponding value of Saint Venant Kirchhoff and 28 \% higher than the linear elastic models (Fig. \ref{fig:solidwall-aaa1-max-von-mises-stress}). Similarly, maximum first principal stress for Raghavan-Vorp model ($\approx 3.90 \times 10^5$ Pa) is 11.4 \% and  18.2 \% higher than the Saint Venant Kirchhoff and linear elastic models (Fig. \ref{fig:solidwall-aaa1-max-principal-stress}). It is important to note that while maximum
$\sigma_{eq}$ and $\sigma_1$ is higher for Saint Venant Kirchhoff model as compared to linear elastic case, it is the otherwise for $\norm{\vb*{u}}_{max}$.

%\begin{figure}[h!]
%	\centering
%	\includegraphics[width=0.65\linewidth]{solidWall_AAA1_max_von_mises_stress_all_models}
%	\caption{Variation of maximum von Mises stress in AAA1 wall over a cardiac cycle for linear, Saint Venant Kirchhoff and Raghavan-Vorp elastic models}
%	\label{fig:solidwall-aaa1-max-von-mises-stress}
%\end{figure}
%
%\begin{figure}[h!]
%	\centering
%	\includegraphics[width=0.65\linewidth]{solidWall_AAA1_max_principal_stress_all_models}
%	\caption{Variation of maximum principal stress in AAA1 wall over a cardiac cycle for linear, Saint Venant Kirchhoff and Raghavan-Vorp elastic models}
%	\label{fig:solidwall-aaa1-max-principal-stress}
%\end{figure}

\subsubsection{Influence of DHr}
\begin{figure}[h!]
\centering
\includegraphics[width=0.5\linewidth]{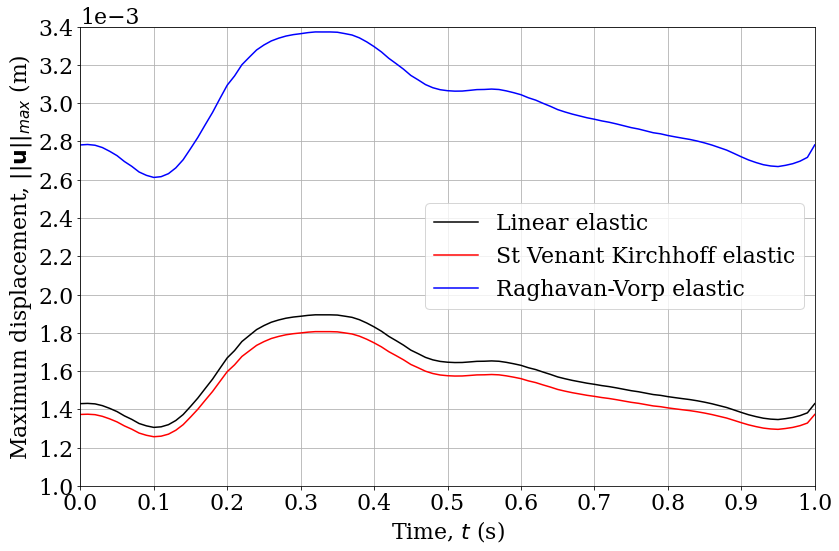}
\caption{Variation of maximum total displacement in AAA2 wall over a cardiac cycle for linear, Saint Venant Kirchhoff and Raghavan-Vorp elastic models}
\label{fig:solidwall-aaa2-max-displacement}
\end{figure}

\begin{figure}[h!]%
\centering
\subfloat[\centering ]{{\includegraphics[width=0.48\textwidth]{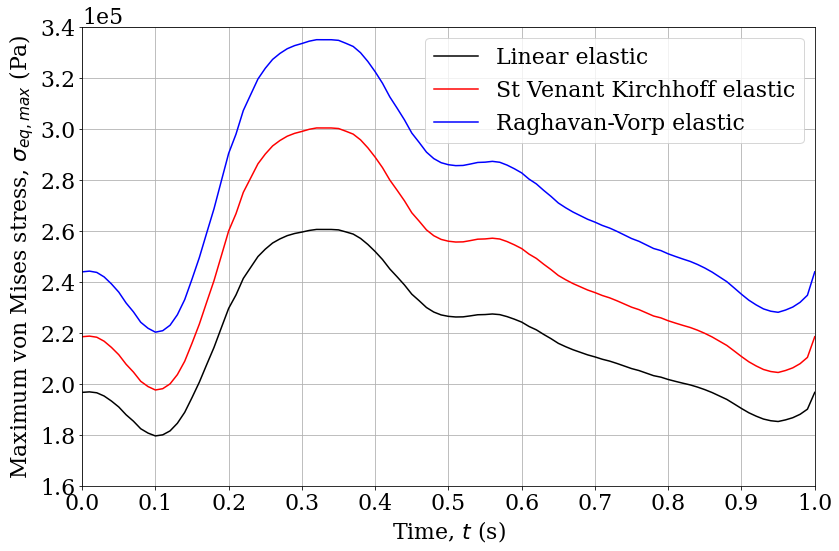} } \label{fig:solidwall-aaa2-max-von-mises-stress}}%
\qquad
\subfloat[\centering ]{{\includegraphics[width=0.48\textwidth]{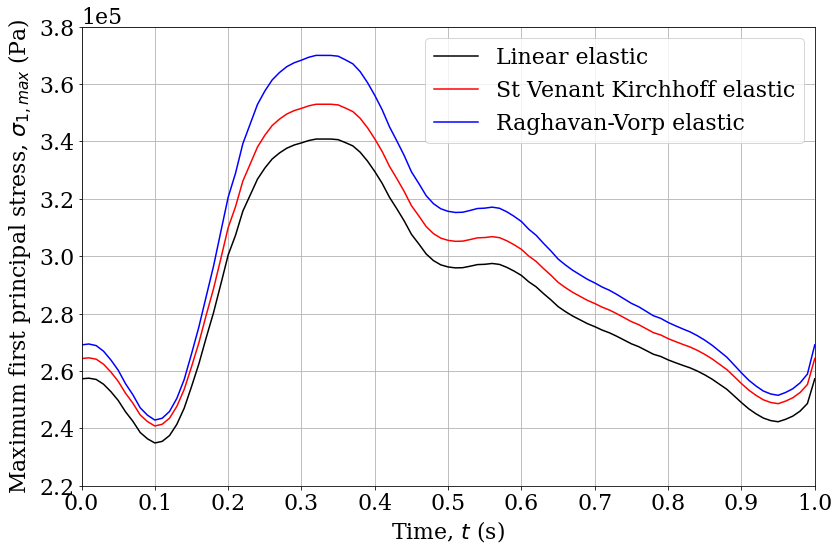} } \label{fig:solidwall-aaa2-max-principal-stress}}%
\caption{Variation of (a) maximum principal stress and (b) maximum von Mises stress in AAA2 wall over a cardiac cycle for linear, Saint Venant Kirchhoff and Raghavan-Vorp elastic models}%
\label{fig:solidwall-aaa2-max-von-mises-principal-stress}%
\end{figure}

Since we have already observed how material model influences the wall mechanics, we shall now consider the results when the AAA1 and AAA2 walls are modeled using the more realistic Raghavan-Vorp elastic model. $\norm{\vb*{u}}_{max}$ of the AAA2 wall (value is $\approx 3.4$ mm) is about 13.3 \% higher than that of AAA1 wall (value is $\approx 3.0$ mm) (see Fig. \ref{fig:solidwall-aaa2-max-displacement}). Maximum $\sigma_{eq}$ in AAA1 wall ($\approx 3.60 \times 10^5$ Pa) is 7.5 \% higher than the corresponding value of AAA2 wall (Fig. \ref{fig:solidwall-aaa2-max-von-mises-stress}). Likewise, maximum $\sigma_{1}$ of AAA1 case ($\approx 3.90 \times 10^5$ Pa) is 5.4 \% higher than that of AAA2 (Fig. \ref{fig:solidwall-aaa2-max-principal-stress}). 

%\begin{figure}[h!]
%	\centering
%	\includegraphics[width=0.65\linewidth]{solidWall_AAA2_max_von_mises_stress_all_models}
%	\caption{Variation of maximum von Mises stress in AAA2 wall over a cardiac cycle for linear, Saint Venant Kirchhoff and Raghavan-Vorp elastic models}
%	\label{fig:solidwall-aaa2-max-von-mises-stress}
%\end{figure}
%
%\begin{figure}[h!]
%	\centering
%	\includegraphics[width=0.65\linewidth]{solidWall_AAA2_max_principal_stress_all_models}
%	\caption{Variation of maximum principal stress in AAA2 wall over a cardiac cycle for linear, Saint Venant Kirchhoff and Raghavan-Vorp elastic models}
%	\label{fig:solidwall-aaa2-max-principal-stress}
%\end{figure}

\cleardoublepage
%%%%%% CONCLUSION 
\section{Conclusion}\label{chap:conclusion}  
%Abdominal aortic aneurysms (AAAs) are local dilatations in the abdominal aorta in the infrarenal segment which occur as a result of weakening of arterial wall due to various predisposing factors, namely intraluminal thrombus, old age, male gender, smoking and hypertension. They are usually asymptomatic until rupture. While traditionally, surgeons tend to use a rupture risk criterion based on maximum transverse diameter of 5-5.5 cm, a better criterion would be the one based on hemodynamic parameters, for example, wall shear stress (WSS), and maximum aneurysmal wall stress which are determined using numerical simulations. 

We have investigated the role of hemodynamic parameters ($TAWSS, WSSG_S, OSI$) and wall stress ($\sigma_{eq}$) in assessing rupture risk of AAAs with respect to the geometric shape index DHr. We draw the following conclusions:
\begin{itemize}
	\item Blood, modeled as a Newtonian fluid, overpredicts WSS and time-averaged WSS ($TAWSS$) throughout the aneurysm sac, primarily in the distal end, and has lower WSS fluctuations leading to lower oscillatory shear index ($OSI$) as compared to non-Newtonian Carreau-Yasuda fluid.
	
	\item For two AAA geometries, AAA1 and AAA2 of different DHr (AAA1 with higher DHr), the flow inside AAA1 forms larger recirculation regions and has richer vortex dynamics than AAA2. On comparing AAA1 and AAA2, we found that AAA1 has lower $TAWSS$ and higher $OSI$ in its proximal end alongwith higher WSS spatial gradient $WSSG_S$ than AAA2. 
	
	\item From AAA wall mechanics simulations, we observe that for a particular AAA, linear elastic and Saint Venant Kirchhoff elastic models underpredict peak wall stress and also the maximum wall displacement as compared to Raghavan-Vorp elastic model. Saint Venant Kirchhoff elastic model gives higher peak wall stress but a lower maximum wall displacement as compared to the linear elastic model. In every case, the peak wall stress is in the midway of the ends of the aneurysm sac and the central bulge location. The peak wall stress of AAA1 is about 8 \% higher than that of AAA2. suggesting that AAA of higher DHr is at a greater risk of rupture. 
	
	\item The analysis show that hemodynamics simulations point to an increased susceptibility to thrombus deposition in the case of AAA1 which is a major predisposing factor for AAA rupture. Overall, DHr is an important determinant for rupture risk prediction.   
\end{itemize}

% Title, author(s), affiliation(s), abstract, text, conclusion, supplementary material section, acknowledgments, author declarations section (conflict of interest, ethics approval, and author contributions), data availability statement, appendixes (if any), and references.

\begin{acknowledgments}
The authors are thankful to Indian Institute Science, Bangalore for fellowship.
\end{acknowledgments}

\section*{Author Declarations}
\subsection*{Conflict of Interest}
The authors have no conflicts to disclose.
\subsection*{Author Contributions}
\textbf{G R Krishna Chand Avatar}: Conceptualization; Data curation;
Formal analysis; Methodology;
 Software; Validation; Visualization; Writing – original draft, review \& editing.
\textbf{Chinika Dangi}: Data curation; 
Software; Validation; Writing – review \& editing.
\textbf{Puneet Pushkar}: Conceptualization; Formal analysis;
Methodology; Software;
Writing – review \& editing.

\section*{Data Availability Statement}
The data that support the findings of this study are available from the corresponding author upon reasonable request.

\bibliography{PoF_AAA_Manuscript}

\end{document}